\documentclass[english,aps,prb, preprint,superscriptaddress, group address,amsmath]{revtex4-2}
\usepackage{fancyhdr,graphicx}
\usepackage[table,xcdraw]{xcolor}
\usepackage[normalem]{ulem}
\useunder{\uline}{\ul}{}
\usepackage[T1]{fontenc}
\usepackage{chemformula} 
\usepackage[utf8x]{inputenc}
\usepackage{gensymb}
\usepackage{times}
\usepackage{color}
\usepackage{amsfonts}
\usepackage{amssymb}
\usepackage[colorlinks,bookmarks=false,citecolor=blue,linkcolor=red,urlcolor=blue]{hyperref}
\input pdfcolor.tex
\usepackage{colortbl,amsthm,amsmath,amssymb,txfonts}
\usepackage{graphicx}
\usepackage{epsfig,color}
\usepackage{verbatim}
\usepackage{dcolumn}
\usepackage[normalem]{ulem}
\usepackage{array}
\usepackage{booktabs}
\usepackage{bm}
\usepackage{xcolor}

\usepackage{orcidlink}
\AtBeginDocument{\pagenumbering{arabic}}
\usepackage{romannum}

\usepackage{graphicx}

\begin{document}

\title{Tunable Band Inversion in Trilayer Graphene}

\author{Harsimran Kaur Mann\orcidlink{0000-0001-7369-550X}$^{1\dagger}$, Simrandeep Kaur\orcidlink{0000-0002-1460-0686}$^{1\dagger}$, Safil Mullick$^1$, Priya Tiwari$^2$, Kenji Watanabe$^3$, Takashi Taniguchi$^4$, Aveek Bid\orcidlink{0000-0002-2378-7980}$^{\ast}$ }

\affiliation{Department of Physics, Indian Institute of Science, Bangalore 560012, India\\
$^2$ Braun Center for Submicron Research, Department of Condensed Matter Physics, Weizmann Institute of Science, Rehovot, Israel\\
$^3$ Research Center for Electronic and Optical Materials, National Institute for Materials Science, 1-1 Namiki, Tsukuba 305-0044, Japan\\
$^4$ Research Center for Materials Nanoarchitectonics, National Institute for Materials Science, 1-1 Namiki, Tsukuba 305-0044, Japan\\
$^\dagger$  These authors contributed equally\\
$^\ast$email: aveek@iisc.ac.in}

\begin{abstract}

Displacement field control of elecronic bands in low-dimensional systems is a promising route toward engineering emergent quantum phases. Here, we report displacement-field-induced band inversion and modulation of the Berry phase of low-energy quasi particles in high-mobility Bernal-stacked trilayer graphene (TLG). Using quantum oscillations, we track the evolution of the Fermi surface and topological properties of Dirac-like “gully” bands that emerge under a finite interlayer potential. We observe a striking sequence of transitions: at low displacement field $D$, the gullies are characterized by a Berry phase of $2\pi$ and large effective mass, indicating massive fermions. As $D$ increases, the Berry phase abruptly shifts to $\pi$ and the effective mass reaches a minimum, signaling the onset of massless Dirac behavior. At higher $D$, the Berry phase returns to $2\pi$, and the effective mass increases again, consistent with a band inversion. These findings demonstrate a rare, reversible topological phase transition—massive → massless → massive—driven entirely by an external displacement field. Despite robust theoretical predictions [\textit{Phys. Rev. B} \textbf{87}, 085424 (2013), \textit{Phys. Rev. B} \textbf{87}, 115422 (2013), and \textit{Phys. Rev. B} \textbf{101}, 245411 (2020)], this evolution of the band topology had escaped experimental detection.  Our results establish TLG as a tunable platform for nanoscale control of band topology. They establish a means to tune between massive and Dirac-like dispersions dynamically providing a foundation for exploring field-switchable topological phenomena in layered 2D systems.

\end{abstract}
\maketitle


\section{INTRODUCTION}

The concept of the Berry phase $\Phi_B$ \cite{doi:10.1098/rspa.1984.0023} pervades virtually every branch of modern quantum condensed matter physics, driving a diverse range of quantum phenomena, including orbital magnetization, electric polarization, and anomalous transport effects~\cite{RevModPhys.82.1959, Manisfestationofbp,PhysRevLett.93.166402}. It is a powerful theoretical tool, allowing physicists to link geometric considerations in quantum wave functions to tangible, measurable properties of materials~\cite{2koshino_trigonal_2009,3zhang_experimental_2005,4zhang_experimental_2011,5novoselov_unconventional_2006,6park_berry_2011,7mikitik_electron_2008}. It is also closely related to topological invariants, such as the Chern number or $\mathcal{Z}_2$ invariant, which characterize topological phases such as topological insulators, Weyl semi-metals, and other exotic quantum phases~\cite{li_topological_2021, PhysRevB.104.205140}. Several topologically non-trivial phases originate from band inversion~\cite{doi:10.1126/science.1148047, Quantumparityhall, PhysRevB.109.045408, PhysRevLett.126.096801, PhysRevLett.102.096801,tiwari_experimental_2022}. The ability to manipulate the Berry phase and control band topology is essential for discovering new materials with robust, topologically protected surface or edge states resistant to disorder and back-scattering.

In this Letter, we show that a perpendicular displacement field $D$ can be used as a tunable control parameter to induce band inversion in ultra-high-quality Bernal-stacked trilayer graphene (TLG). Our study reveals a strong  $D$--tunability of $\Phi_B$ and $m^*$. Specifically, at low $D$, the low-energy trigonally warped bands (termed Dirac gullies) are massive with a relatively large $m^*$, and $\Phi_B=2\pi$. As $D$ increases, the gullies deepen, and $m^*$ decreases sharply, accompanied by a transition of $\Phi_B$ to $\pi$, signifying a transition to massless Dirac gullies. Intriguingly, at still higher $D$, the effective mass $m^*$ increases again, suggesting a band inversion~\cite{gateinduceddiraccones}, with $\Phi_B$ reverting to $2\pi$.

\begin{figure*}[t]
\centering
     \includegraphics[width=\columnwidth]{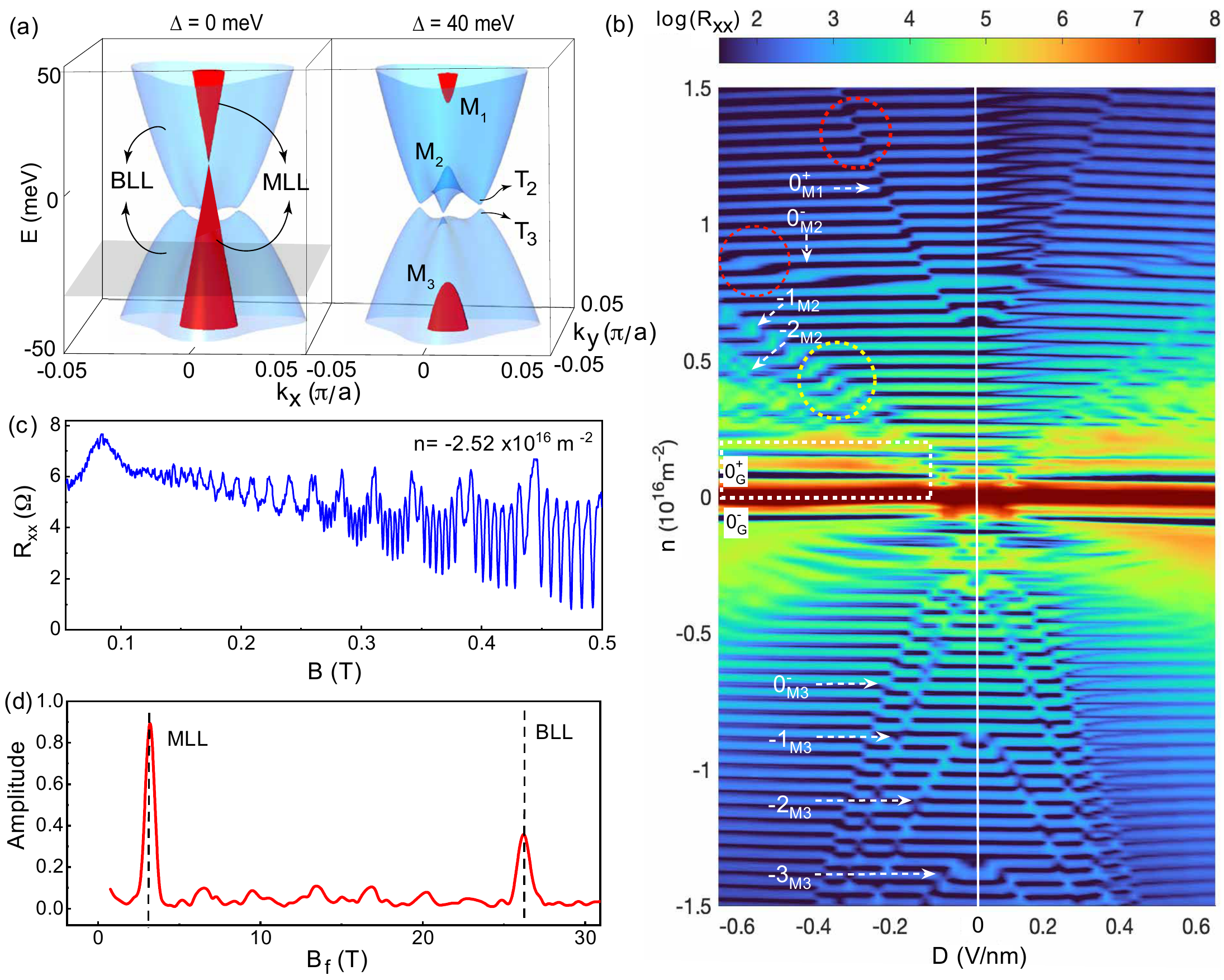}
     \caption{\textbf{Device Characterization}: (a) Calculated band structure of TLG at $\Delta=0$~meV and $\Delta=40$~meV. Here a~$=2.46\times10^{-10}~\mathrm{m^{-2}}$ is carbon-carbon distance. (b) Contour plot of  $R_{xx}$ versus $n$ and $D$ on a log scale. The data were measured at $T=20$~mK and $B=0.5$~T. The labels in the plot refer to the Landau levels. The red dotted circles show a phase shift $\pi$ due to crossings between the BLL Landau level and $0_{M2}^+$ and $0_{M1}^-$ of the valley-polarized MLL Landau levels. The yellow dotted circle marks the region of $2\pi$ phase shift due to crossing between the BLL band and $-1_{M_2}$ of the MLL band. (c) SdH oscillations measured at $n = -2.52\times 10^{16}~\mathrm{m^{-2}}$, $T=20$~mK and $\Delta = 0$~meV. The data were taken along the gray plane marked in (a). (d) Fast-Fourier Transform of the data in (b) showing two primary frequencies corresponding to the MLL and BLL bands.}
\label{fig1}
\end{figure*}

TLG is a multi-band system formed by monolayer-like (MLL) linear bands and bilayer-like (BLL) quadratic bands (Fig.~\ref{fig1}(a)) at zero interlayer potential ($\Delta=0$ meV). These two bands are protected by mirror symmetry from intermixing. Applying a displacement field $D$ perpendicular to the interface breaks the mirror symmetry and hybridizes the MLL and the BLL bands. In the presence of a finite $D$, the MLL band splits into three sections: $M1$, $M2$, and $M3$, as shown in the right panel of Fig.~\ref{fig1}(a). With increasing  $D$, the MLL bands $M1$ and $M3$ shift to higher energies. Importantly, for our purposes, the  $M2$ band merges with the BLL band and develops into a novel set of $C_3$-symmetric Dirac gullies~\cite{PhysRevLett.113.116602,PhysRevB.93.115106,newdiracpoints, GullyQHferro,  gateinduceddiraccones, PhysRevB.86.045407, VARLET201519}. This band modulation leads to novel correlated phases~\cite{Seiler2022, PhysRevLett.134.036301}, interaction-driven nematic phases~ \cite{GullyQHferro, emergentdiracgullies} and multiple Lifshitz transitions~\cite{imagingquantumoscillations, PhysRevLett.113.116602, Nonlocaltlg}.

Fig.~\ref{fig1}(b) is a contour plot of the longitudinal resistance $R_{xx}$ as a function of $n$ and $D$, measured at $B = 0.5$~T. The plot shows crossings between the Landau levels of the MLL band (labeled $M1$, $M2$, and $M3$) and the four-fold degenerate BLL bands. At finite $D$, the $M_2$ band emerges, and the $M_1$ and $M_3$ bands move to higher number densities. Prior studies have extensively investigated Berry phase changes in various regimes of this phase diagram~\cite{imagingquantumoscillations}. Crossings between the BLL band landau level and $0_{M2}^+$ and $0_{M1}^-$ of the valley polarized MLL band landau levels result in a $\pi$ phase shift in the region marked by the red dotted circles in Fig.~\ref{fig1}(b).
Crossing between the landau levels of BLL band and $-1_{M_2}$ of the MLL band results in a phase shift of $2\pi$ (yellow dotted circle in Fig.~\ref{fig1}(b)). Furthermore, the Berry phase of the massive band changes from $\pi$ to $-\pi$ as the Fermi level is tuned from below the band gap to above the band gap of the Dirac like band at $D=0$~V/nm (along the white solid line in Fig.~\ref{fig1}(b))~\cite{Nontrivialqhbeeryphase}.

In all previous experimental studies, a study of $\phi_B$ and band mass evolution in the regime characterized by low carrier density and high-$D$ (region bounded by the white dotted rectangle in Fig.~\ref{fig1}(b)), where theoretical models predict the emergence of electronic `gullies', has remained unexplored. These gullies are expected to host Dirac-like band topology, each carrying a Berry phase of $\pi$~\cite{2koshino_trigonal_2009, gateinduceddiraccones}. Furthermore, the effective mass $ m^*$ of these Dirac-like features is predicted to be $D$-tunable, with a band inversion expected beyond a critical displacement field~\cite{gateinduceddiraccones, GullyQHferro, newdiracpoints}. In this Letter, we report the first experimental realization of these theoretically anticipated features.\\

\section{RESULTS AND DISCUSSION}

\subsection{Probing $D$-dependence of band structure using SdH oscillations} The measurements were performed on a high-mobility hexagonal boron nitride (hBN) encapsulated TLG device with top and back graphite gates (Appendix-A)~\cite{kaur_universality_2024}. The dual graphite gate configuration is used for simultaneous tuning of the perpendicular displacement field $D=[(C_{bg}V_{bg}-C_{tg}V_{tg})/{2\epsilon_{0}}+D_{0}]$ and charge carrier density $n = [(C_{bg}V_{bg} + C_{tg}V_{tg})/e + n_{0}]$ independently. Here $C_{bg}$ ($C_{tg}$) is the back-gate (top-gate) capacitance, and $V_{bg}$ ($V_{tg}$) is the back-gate (top-gate) voltage.  $D_{0}$ and $n_{0}$ are the graphene's residual displacement field and number density due to impurities. All the measurements were performed using standard low-frequency lock-in detection techniques in a dilution refrigerator with a base temperature of $20$~mK.

The measured SdH oscillations at $n = -2.52\times 10^{16}~\mathrm{m^{-2}}$ at $D=0~\mathrm{V/nm}$ are presented Fig.~\ref{fig1}(c); the data are taken along the gray plane in Fig.~\ref{fig1}(a).  The beating pattern in the oscillations arises from the coexistence of two distinct Fermi surface areas. Recall that at fixed chemical potential (equivalently fixed $n$), the amplitude of SdH oscillations $\Delta R_{xx}\propto\mathrm{cos}[2\pi({B_f}/{B}+\gamma)]$ with $B_f=\hbar S/({2\pi e})=nh/(ge)$ and $\gamma=({\Phi_B}/{2\pi}-{1}/{2})$~~\cite{PhysRevB.96.205410,1novoselov_two-dimensional_2005}. Here, $R_{xx}$ is the longitudinal resistance, $S$ is the area enclosed by the Fermi contour, $g$ is the degeneracy of Landau levels, and $\Phi_B$ is the Berry phase. The Fast Fourier transform (FFT)  of oscillations shows two primary frequencies (Fig.~\ref{fig1}(d)). The lower (higher) frequency corresponds to the MLL (BLL) band with a smaller (larger) Fermi contour area. By systematically tracking these frequencies at different $n$ and $D$, we map out the complex band structure of ABA TLG and its evolution with $D$.

\begin{figure}[h!]
\includegraphics[width=\columnwidth]{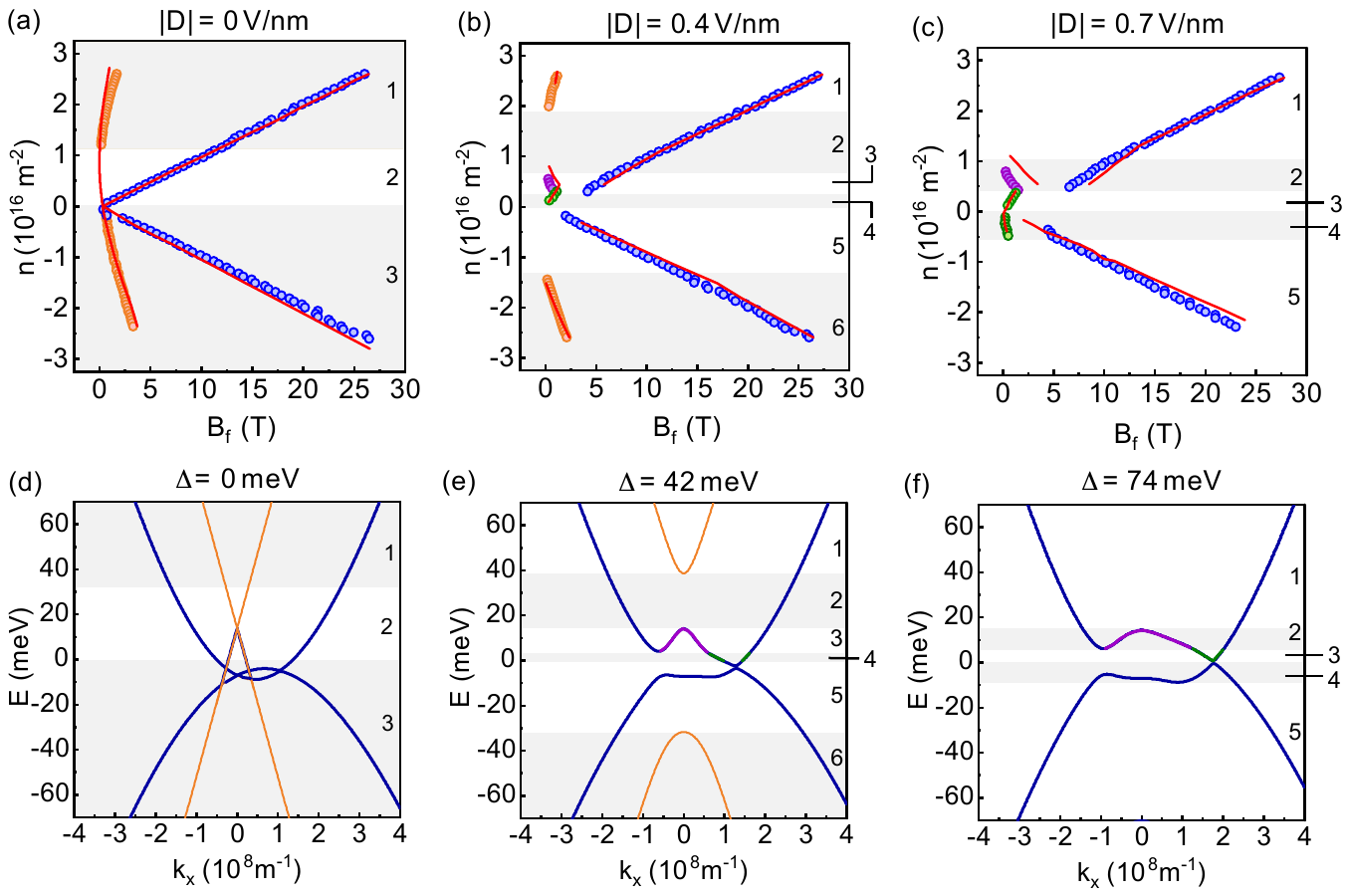}
\caption{\textbf{D-dependent band structure}: Number density $n$ versus SdH frequency $B_f$  obtained experimentally for constant $D$ (a) $|D|=0~\mathrm{V/nm}$ (b) $|D|=~0.4\mathrm{V/nm}$ and (c) $|D|=0.7~\mathrm{V/nm}$. The open circles show the experimentally obtained $B_f$; Blue- BLL band , orange- MLL band $M
_1$ and $M_3$, violet- MLL band $M_2$  and green- Gully. Red solid lines show the theoretically calculated $B_f$ as  function of $n$ calculated from the  $E-k$ diagrams shown in  (d) $|D|=0~\mathrm{V/nm}$ (e) $|D|=0.4~\mathrm{V/nm}$ and (f) $|D|=0.7~\mathrm{V/nm}$. Regions $1-6$ show the correspondence between the $n-B_f$ and $E-k$ plots.}
\label{fig2}
\end{figure}

Fig.~\ref{fig2}(a-c) shows $B_f$ versus $n$ plots measured at $D=0~\mathrm{V/nm}$, $0.4~\mathrm{V/nm}$ and $0.7~\mathrm{V/nm}$, respectively. The calculated band structure at these values of displacement fields $D$ is plotted in Fig.~\ref{fig2}(d-f). The data acquired at $D=0~\mathrm{V/nm}$ can be divided into three distinct regions; these are marked in Fig.~\ref{fig2}(a) and (d). In regions $1$ and $3$, we identify the Monolayer-like (MLL) band (orange open circles) and the Bilayer-like(BLL) band (blue open circles). In region $2$, near the Dirac point of the MLL band, the fermi surface area $S$ (and consequently $B_f$) becomes immeasurably small. We are thus left with only the value of $B_f$ for the BLL band in this region.

At $D=0.4~\mathrm{V/nm}$, the energy spectrum becomes markedly more complex as seen from Fig.~\ref{fig2}(b) and Fig.~\ref{fig2}(e). It is now convenient to divide the data into six distinct regions. As discussed in the introduction in the main paper, at a finite-$D$, the MLL band is split into three parts; we identify all three segments of the MLL band from our data along with the Dirac gullies. In region~1, we find  $M_1$ (open orange circles) and the BLL (open blue circles) bands. Region~2 lies in the gap of the MLL band (Fig.~\ref{fig2}(e)); this explains the appearance of only a single $B_f$ in this range of number density (energy) originating from the BLL band (Fig.~\ref{fig2}(b)). Region~3 hosts the $M_2$ section of the MLL band (open violet circles in Fig.~\ref{fig2}(b)) and the tail of the BLL band (open blue circles in Fig.~\ref{fig2}(b)).  At very low number densities (region~4), the Dirac gullies emerge (Fig.~\ref{fig2}(e)) -- their oscillation frequencies are plotted as open green circles in Fig.~\ref{fig2}(b). On the hole side, region~5 contains only the BLL band while region~$6$ has both the $M_3$ (open orange circles) and the BLL (open blue circles) bands.

Similar features are observed for $D=0.7~\mathrm{V/nm}$ (Fig.~\ref{fig2}(c) and Fig.~\ref{fig2}(f)). At this high value of $D$, the $M_1$ and $M_3$ bands move to number densities beyond our measurement range. Regions~$1$ and $5$ thus have only the BLL bands. Regions~$2$ has the $M_2$ band and the BLL band. The region~$3$ where the Dirac Gully band forms widens in range as compared to that at $D=0.4~\mathrm{V/nm}$. Region~$4$ shows the Dirac Gully on the hole side.

The theoretically calculated $n$ versus $B_f$ is shown by the red solid line in Fig.~\ref{fig2}(a-c). It was calculated from the theoretical band structures (Fig.~\ref{fig2}(d-f)) by calculating the area of the Fermi surface ($S$) at constant energy and using the relation $B_f= \hbar S/2\pi e$. The striking resemblance of the experimental data to the theoretical plot shows the accuracy of the measurement and sensitivity to different Fermi surfaces in the band.

\subsection{Detailed SdH map in the Gully region}

\begin{figure*}[t]
    \centering
    \includegraphics[width=\columnwidth]{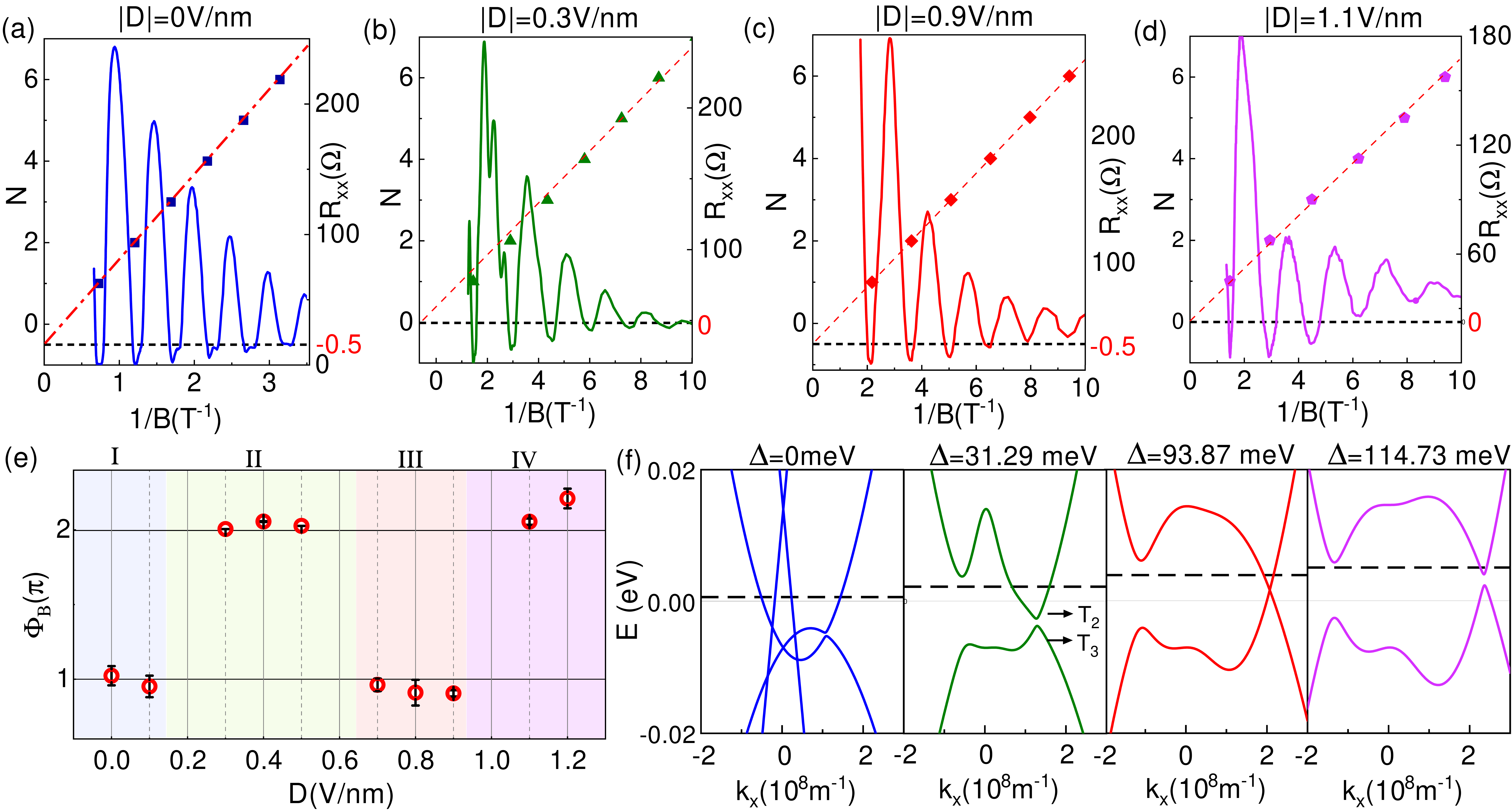}
    \caption{\textbf{Dependence of $\nu$ on $D$} : (a) Contour map of $R_{xx}$ versus $D$ and $\nu$ measured at a fixed  $n = 1.8\times 10^{15}~\mathrm{m^{-2}}$ and $T = 20$~mK. (b) Calculated Fermi contour and (c) FFT of SdH oscillations for $D$ values corresponding to region~1 ($\Delta = 0$~meV, $D=0$~V/nm), region~2 ($\Delta = 31$~meV, $|D|=0.3$~V/nm), region~3 ($\Delta = 94$~meV, $|D| = 0.9$~V/nm) and region~4 ($\Delta = 115$~meV, $|D| = 1.1$~V/nm).}
    \label{fig3}
\end{figure*}

A map of the SdH oscillations as a function of filling factor $\nu$ and $D$ measured at $n = 1.8\times 10^{15}~\mathrm{m^{-2}}$ is presented in Fig.~\ref{fig3}(a). We identify four distinct $D$-regions with unique Landau level sequences. For low displacement fields $0\leq D \leq 0.15~\mathrm{V/nm}$ (labeled  \textbf{region~\Romannum{1}}), the $R_{xx}$ minima appear at $\nu=14,18,22...$, following the sequence $\nu=4\times(N_B+1/2)$, where $N_B$ is the Landau level index of BLL (the factor $1/2$ arises from the underlying zeroth Landau levels of MLL band \cite{Nontrivialqhbeeryphase}). As $D$ increases, BLL bands deform into three different gullies (\textbf{region~\Romannum{2}},  $0.15~\mathrm{V/nm}\leq D \leq 0.65~\mathrm{V/nm}$). The Landau level sequence in this range of $D$  changes abruptly to  $\nu=12\times N_G = 12, 24, 36, ...$, where $N_G$ is the Landau level index of gullies. Interestingly, as we increase $D$ further (\textbf{region~\Romannum{3}}, $0.65~\mathrm{V/nm}\leq D \leq 0.9~\mathrm{V/nm}$), we observe prominent minima at $\nu=18,30,42$, changing the Landau level sequence to $\nu=12\times (N_G+1/2)$. At much higher $D$, the Landau level sequence reverts to $\nu=12\times N_G$ (\textbf{region~\Romannum{4}}, $D \geq 0.9~\mathrm{V/nm}$).

\begin{figure*}[t]
    \centering
    \includegraphics[width=\columnwidth]{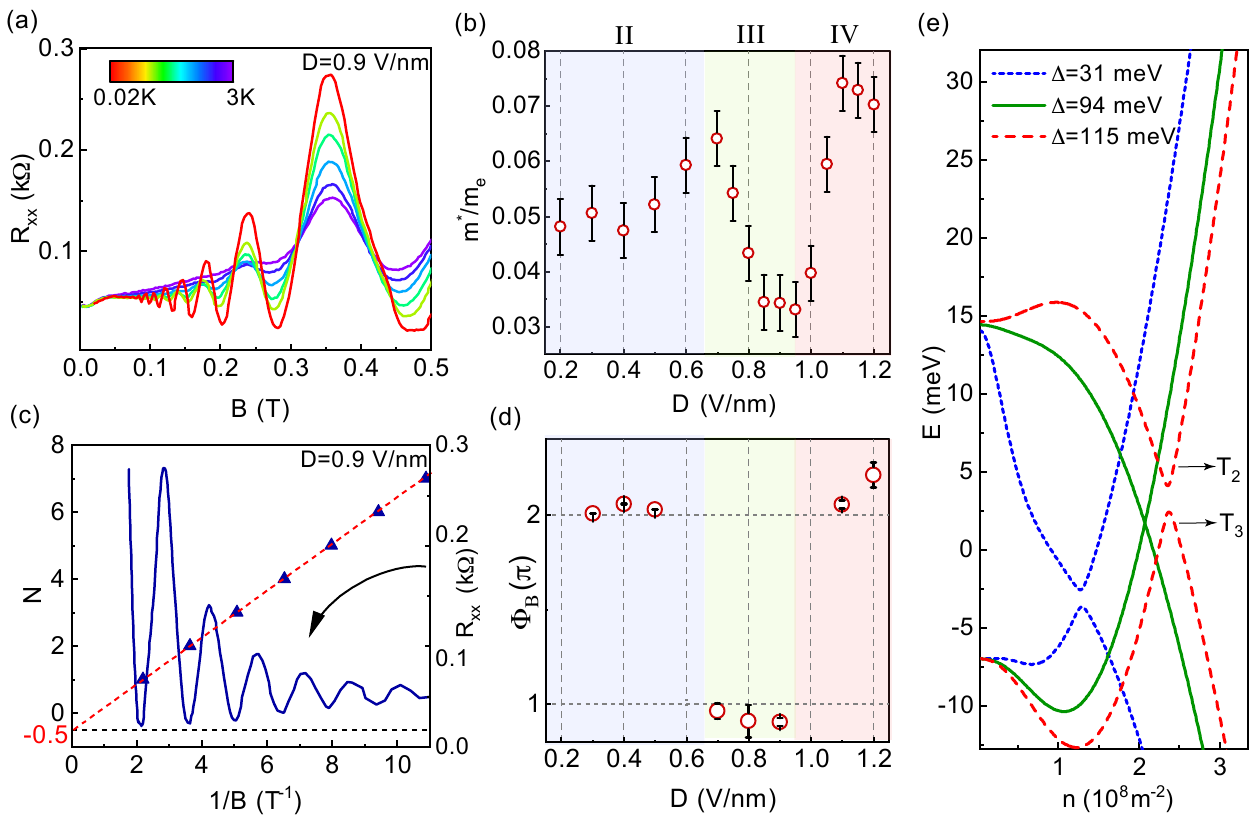}
    \caption{\textbf{Estimation of effective mass and Berry phase of Dirac gullies:} (a) Plots of the $R_{xx}$ as a function of $B$ measured for $D=0.9$~V/nm over the temperature range $0.02$ K$\leq T \leq 3$ K. (b) Effective mass $m^*$ (scaled by the mass of free electron $m_e$) as a function of $D$ (red open circles). The error bars result from the scatter in the value of $m^*$ when calculated from the different peaks in (a). The Roman numerals on the top indicate the four distinct regimes of $D$. (c) Plots of the Landau level index $N$ (left axis) and $R_{xx}$ (right axis) as a function of $1/B$ for $D=0.9$~V/nm. The scatter plot shows $N$ vs $1/B$ (blue triangles). The dotted red line linearly fits the $N$ versus $1/B$ data; the y-intercepts give $\Phi_B$. (d) $\Phi_B$ as a function of $D$. (e) Calculated band structure of $T_2$ and $T_3$ gullies for four representative values of $D$ in region~2 ($\Delta = 31$~meV, $|D|=0.3$~V/nm), region~3 ($\Delta = 94$~meV, $|D| = 0.9$~V/nm) and region~4 ($\Delta = 115$~meV, $|D| = 1.1$~V/nm).}
    \label{fig4}
\end{figure*}

Fig.~\ref{fig3}(b) shows the Fermi surface contours at different interlayer potential $\Delta$, calculated using tight-binding based on the Slonczewski-Weiss-McClure model \cite{Dresselhaus01012002, PhysRev.109.272, MCCLURE1969425, GullyQHferro}. At low-$D$ (\textbf{region~\Romannum{1}}), the perturbation to the pristine band structure is minimal, and one can still distinguish the MLL and BLL bands. We find two Fermi contours; the inner red circular contour arises from the MLL band, while the outer trigonally warped contour is from the BLL band.  Consequently, at low-$D$ and low-$n$, one observes the characteristic $4N_B+2$ Landau level progression with $4N_B$ Landau levels coming from the BLL band and the $2$ additional Landau levels from the zeroth Landau level of the MLL band \cite{Nontrivialqhbeeryphase}. The FFT spectrum of the SdH oscillations reveals a single frequency at $2.03$~T (Fig.~\ref{fig3}(c), first panel), which corresponds to the BLL band; the frequency of the MLL band ($B_f\leq 0.4$~T) is too small to be observed in low-$B$ SdH measurements.

We now turn to the results at higher $D$. As noted earlier, a large displacement field induces significant mixing between the monolayer-like (MLL) and bilayer-like (BLL) bands, blurring their distinct identities. However, we continue to label these bands as MLL and BLL for clarity and consistency, even at finite $D$. At $D = 0.3$ V/nm ($\Delta = 31$ meV), the Dirac gullies appear, marking the onset of \textbf{region~\Romannum{2}}, where the low-energy BLL band fragments into three disconnected Fermi pockets (Fig.~\ref{fig3}(b), second panel). This reconstruction changes the Landau level degeneracy to 12 ($3$ gullies, each with $2$-fold spin and $2$-fold valley degeneracy). Notably, the MLL band remains at much higher energy and is unoccupied at this $D$.

Because the charge carrier density is held constant at $n = 1.8 \times 10^{15}\mathrm{m^{-2}}$, the area of each Fermi pocket (and the corresponding $B_f$) is expected to reduce to one-third of its value in \textbf{region~\Romannum{1}} (recall that $B_f \propto n/g$). This expectation is confirmed by the FFT spectrum comparison in Fig.\ref{fig3}(c) between \textbf{regions \Romannum{1}} and \textbf{\Romannum{2}}. In \textbf{regions \Romannum{3}} and \textbf{\Romannum{4}}, the gullies become further separated; however, since both $n = 1.8 \times 10^{15}\mathrm{m^{-2}}$ and the Landau level degeneracy $g = 4$ remain unchanged from \textbf{region~\Romannum{2}}, the measured $B_f$ (and thus the Fermi pocket area) remains constant within experimental uncertainty.

\subsection{Tuning the effective mass and Berry phase with $D$}

The abrupt changes in the Landau level sequence noted in Fig.~\ref{fig3}(a) indicate $D$-induced transitions in band topology. We extract the $D$-dependent effective mass $m^*$ and Berry phase $\Phi_B$ to quantify these transitions. Representative plots of the temperature-dependent SdH oscillations at $D=0.9~\mathrm{V/nm}$, measured over the temperature range $20$~mK--$3$~K, are plotted in Fig.~\ref{fig4}(a). The effective mass $m^*$ is extracted by fitting the maxima of the amplitude of the $R_{xx}$ oscillations at a fixed $B$ to the equation~\cite{PhysRevB.96.205410}:

\begin{equation}
 \Delta R_{xx} \propto \frac{\lambda}{\mathrm{sinh} \lambda }  \mathrm{cos}\left( 2\pi\left( \frac{B_f}{B} +\gamma \right)\right)
 \label{eq1}
 \end{equation}
 with $\lambda=(2\pi^2 k_B T m^*)/(\hbar eB)$ (Appendix -- B).
 Fig.~\ref{fig4}(b) shows the extracted effective mass for the three gully regions (red open circles).

Concurrently, we track the evolution of the Berry phase across these transitions. Fig.~\ref{fig4}(c) show plots of $R_{xx}$ (right axis) and the Landau level index $N$ corresponding to minima in $R_{xx}$ (left axis)  as a function of $1/B$  for $D=0.9~\mathrm{V/nm}$. The y-axis intercept of the linear fit to the data (red dotted line in Fig.~\ref{fig4}(c)) gives $\Phi_B$ in units of $\pi$ (Appendix -- D).
Fig.~\ref{fig4}(d) shows the extracted $\Phi_B$ for the three gully regions.

The evolution of effective mass and $\Phi_B$ with $D$ in the gully region is non-trivial. As noted earlier, the gully regions begin to develop in \textbf{region~\Romannum{2}} with $m^{*} \sim 0.05 \pm 0.005m_e$ (Fig.~\ref{fig4}(b) -- blue shaded region) and $\Phi_B = 2\pi$ (Fig.~\ref{fig4}(d) -- blue shaded region). Thus, in this regime of the phase space, the low-energy charge carriers are \textit{chiral massive
 particles}. With increasing $D$, the value of  $m^{*}$ passes through a sharp minimum at $D_c = 0.9$~V/nm (Fig.~\ref{fig4}(b) -- green shaded region). Concurrently, the Berry phase jumps abruptly to $\Phi_B = \pi$ (Fig.~\ref{fig4}(d) -- green shaded region), indicating that in this range of $D$ values, the charge carriers revert to being \textit{chiral massless Dirac particles}.
 With the further enhancement of $D$; the effective mass again becomes finite (Fig.~\ref{fig4}(b) -- pink shaded region) and $\phi_B$ jumps back to $2\pi$ (Fig.~\ref{fig4}(d) -- pink shaded region) establishing that the change carriers in this regime are \textit{chiral massive
 particles}.

The $D$-dependence of $m^*$ thus establishes a massive $\rightarrow$ massless $\rightarrow$ massive topological phase transition of the Dirac gullies. This conclusion is supported by the concurrent evolution of the measured Berry phase in the sequence $2\pi \rightarrow \pi \rightarrow 2\pi$. This reversible transformation of the charge carriers' chirality is the central finding of this study, providing a unique opportunity to dynamically switch between massive and Dirac-like dispersions within the single band of a given material system.

The observed evolution of $\Phi_B$ with $D$ corroborates well with our calculations and previous theoretical predictions~\cite{gateinduceddiraccones, GullyQHferro,newdiracpoints}. Fig.~\ref{fig4}(e) shows the calculated band structure at three representative values of $D$ corresponding to the three regions. In \textbf{region~\Romannum{2}}, the low energy dispersion has a distorted parabolic shape (blue dotted line in Fig.~\ref{fig4}(e); data plotted for $\Delta = 31$~meV). In contrast, the dispersion relation becomes linear in \textbf{region~\Romannum{3}} (green solid line in Fig.~\ref{fig4}(e); data plotted for $\Delta = 94$~meV). In this region of the phase space, the charge carrier velocity estimated from the measured band mass, $v_F = \hbar\sqrt{n\pi}/m^* = 3\times 10^5~\mathrm{ms^{-1}}$ matches well with the band velocity calculated from the dispersion relation, $v_F = (1/\hbar)(\partial E/\partial k) = 4\times 10^5~\mathrm{ms^{-1}}$.   For $|D| > 0.9$~V/m (\textbf{region~\Romannum{4}}, red dashed-dotted line in Fig.~\ref{fig4}(e); data plotted for $\Delta = 115$~meV), the low-energy spectrum changes back to quasi-parabolic (Appendix-E).

\section{CONCLUSION}

Our measurements reveal that the Berry phase evolution in the gully band of Bernal-stacked trilayer graphene arises from a topological transition intrinsic to the gully band itself, in contrast to the situation at $n=0$~\cite{Nontrivialqhbeeryphase}. Our band structure mapping substantiates this, which shows a single-frequency SdH oscillation, indicating that only one band contributes to the observed Berry phase.
Importantly, the transition sequence $\phi_B: 2\pi \rightarrow \pi \rightarrow 2\pi$ reflects a reversible change in the nature of the charge carriers—from massive to massless and back—consistent with a theoretically predicted band inversion between the $T_2$ and $T_3$ gully bands \cite{gateinduceddiraccones,GullyQHferro,newdiracpoints}. This inversion is expected to occur near an interlayer potential $\Delta \sim 92$~meV \cite{GullyQHferro}, where the effective mass reaches a minimum. Our experimental extraction of effective mass supports this prediction, showing a gap closure around $\Delta = 94$ meV (Fig.~\ref{fig4}(b)).

These findings directly validate a $D$-tunable  topological phase transition within a single band, with the massless Dirac-like behavior of gullies confined to a narrow range of displacement fields near the critical point.  This reversible evolution offers a rare opportunity to dynamically toggle between massive and Dirac-like dispersions within a single band of a 2D material system. The ability to control band topology and Berry phase through electrostatic gating opens up exciting possibilities for engineering topological states and chiral quasiparticles in layered 2D materials, with potential implications for valleytronics and quantum devices.\\

 \section*{Acknowledgment}
The authors acknowledge fruitful discussions with Tanmoy Das. A.B. acknowledges funding from the Department of Science \& Technology FIST program and the U.S. Army DEVCOM Indo-Pacific (Project number: FA5209   22P0166). K.W. and T.T. acknowledge support from JSPS KAKENHI (Grant Numbers 19H05790, 20H00354, and 21H05233).

\clearpage

\section{Appendix}

\subsection{Device Fabrication \label{app:device}}

\begin{figure}[h!]
\includegraphics[width=0.9\columnwidth]{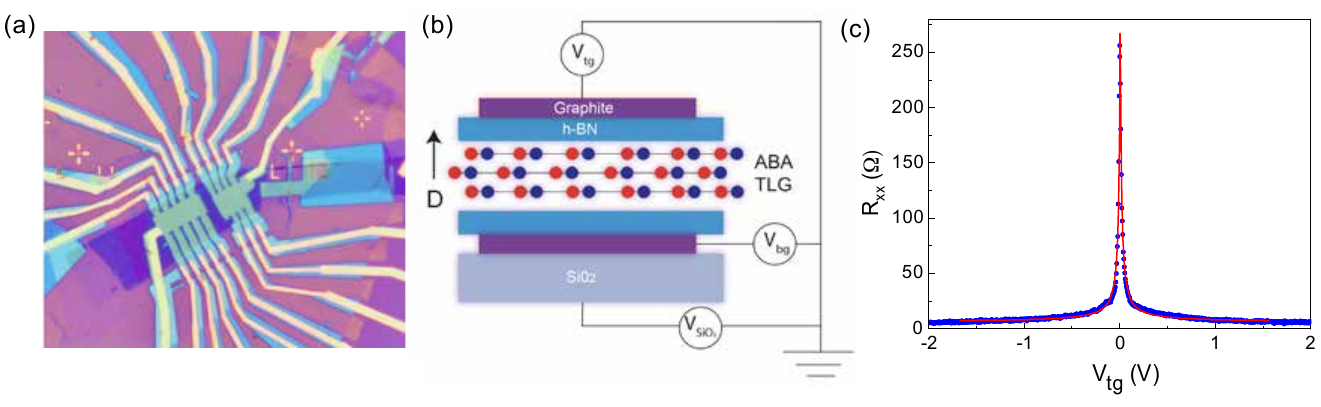}
\caption{\textbf{Device Characterization}:(a) Optical image of dual graphite gated TLG device. (b) Schematic of the device. (c) Measured $R-V_g$ curve (blue circles). The fit (solid red curve) gives mobility $\mu= 62~\mathrm{m^2V^{-1}s^{-1}}$ and the intrinsic carrier concentration induced by charge impurity $n_0=7.81 \times 10^{13}~\mathrm{m^{-2}}$.}
\label{figS1}
\end{figure}

Trilayer graphene (TLG) and hexagonal boron nitride (hBN) flakes were mechanically exfoliated onto $\ch{Si}/\ch{SiO2}$ substrates. The hBN flakes had a thickness of 25-30 nm. The TLG flakes were identified from optical contrast under a microscope and later confirmed using Raman spectroscopy. To make the stack, the flakes are picked up using polycarbonate film (PC film)in the sequential order of graphite/hBN/TLG/hBN/graphite. The stack is then transferred to \ch{Si}/\ch{SiO2} substrate along with PC film, followed by removing the PC residue in chloroform. The heterostructure is then annealed in a vacuum at $300^\circ C$ for 4 hours. Electron beam lithography was used to define the contacts. This was followed by etching with a mixture of $\ch{CHF_3}$ (40 sscm) and $\ch{O_2}$ (10 sscm). The metallization was done with $\ch{Cr/Pd/Au}$ (3~nm/12~nm/55~nm) to form the 1-D contacts with TLG. Fig.~\ref{figS1}(a) shows an optical image of the device. The graphene contacts are doped to higher number densities using $\ch{SiO2}$ gate to avoid p-n junction formation~\cite{kaur_universality_2024}.

To calculate the mobility of the device the measured $R_{xx}$ as a function of number density at $D=0\mathrm{V/nm}$ and $B=0T$ is fit to the equation $R=R_c +\frac{L}{We\mu \sqrt{n^2+{n_0}^2}}$. Here,  $R_c$, $L$, $W$, and $\mu$ are the contact resistance, length, width, and mobility of the device, respectively. From the fit (Fig.~\ref{figS1}(c)), the extracted mobility is $\mu= 62~\mathrm{m^2V^{-1}s^{-1}}$ and the intrinsic carrier concentration induced by charge impurity $n_0=7.81 \times 10^{13}~\mathrm{m^{-2}}$ reflecting the high quality and low impurity levels of the sample.

\subsection{Calculation of effective mass $m^*$ from SdH oscillations}

\begin{figure}[h!]
\includegraphics[width=\columnwidth]{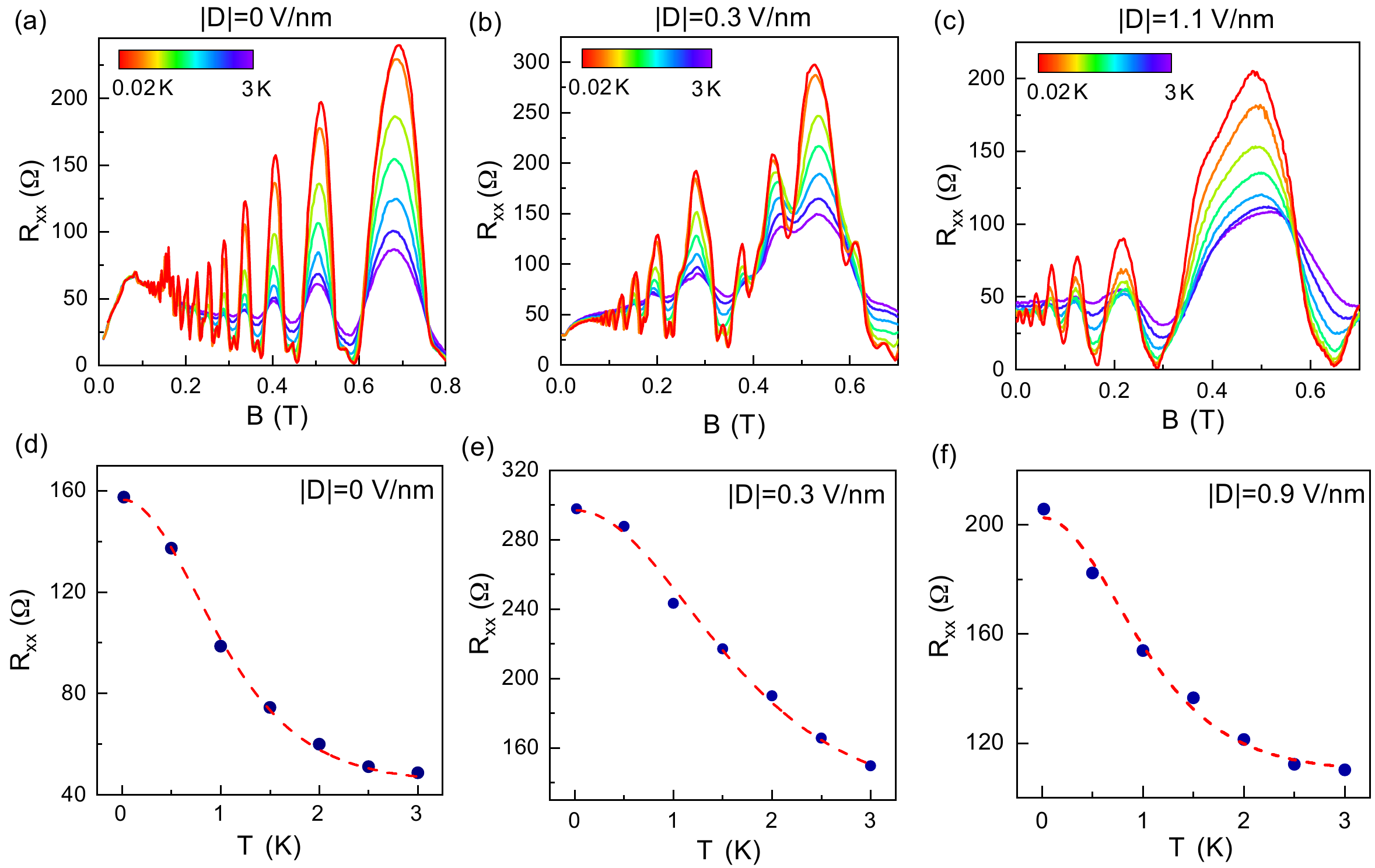}
\caption{\textbf{Calculation of effective mass}: Plots of the $R_{xx}$ as a function of B measured for (a) $|D|=0~\mathrm{V/nm}$, (b) $|D|=0.3~\mathrm{V/nm}$ and (c) $|D|=1.1~\mathrm{V/nm}$ over the temperature range $0.02$~K$\leq T\leq3$~K. (d) Plots of the amplitude of SdH oscillations as a function of $T$ for $|D|=0~\mathrm{V/nm}$, (e) $|D|=0.3~\mathrm{V/nm}$ and (f) $|D|=1.1~\mathrm{V/nm}$. The red dotted lines fit the data points using the Lifshitz-Kosevich relation.}
\label{figS2}
\end{figure}

The effective mass of the Dirac Gully was measured as a function of $D$. Representative plots of the temperature-dependent SdH oscillations at $D=0~\mathrm{V/nm}$, $D=0.3~\mathrm{V/nm}$ and $D=1.1~\mathrm{V/nm}$ measured over the temperature range $20$~mK--$3$~K are plotted in Fig.~\ref{figS2}(a~-~c). The effective mass $m^*$ is extracted by fitting the maxima of the amplitude of the $R_{xx}$ oscillations (blue circles in Fig.~\ref{figS2}(d~-~f)) at a fixed $B$ to the equation~\cite{PhysRevB.96.205410}:
\begin{equation}
\Delta R_{xx} \propto \frac{\lambda}{\mathrm{sinh} \lambda }  \mathrm{cos}\left( 2\pi\left( \frac{B_f}{B} +\gamma \right)\right)
\label{eq2}
exp \left( \frac{-2\pi^2k_BT_D}{\hbar eB}\right)
\end{equation}
with the amplitude of oscillations,
\begin{equation}
 \lambda=\frac{2\pi^2 k_B T m^*}{\hbar eB}
\label{eq3}
\end{equation}
The dotted red lines in Fig.~\ref{figS2}(d~-~f) show the fit to the $R_{xx}$ maxima to Eqn.~\ref{eq2} at different values of  $D$.

\subsection{Calculation of $\nu$ from Landau level Index $N$}

The filling factor $\nu$ can be expressed as a function of $N$ as $\nu=\alpha N+\beta$ where $\alpha$ and $\beta$ are constants. The following table summarizes the observed values of $\alpha$ and $\beta$ for the different bands in this study.

\begin{table}[h!]
    \centering
    \begin{tabular}{m{3cm} m{2cm} m{2cm} m{3cm}}

         &   \textbf{$\alpha$}    &    \textbf{$\beta$ }   &    \textbf{$\nu=\alpha N+\beta$}    \\
         \hline
        BLL band & $4$ & $2$ & $6, 10, 14, 18...$\\

         Massive Gully& $12$ & $0$ & $12, 24, 36, 48....$\\

         Massless Gully& $12$ & $6$ & 18, $30, 42, 54...$ \\

    \end{tabular}
    \caption{$\alpha$ and $\beta$ values corresponding to different bands in this study.}
    \label{tab1}
\end{table}

\subsection{Calculation of Berry Phase $\Phi_B$ from SdH oscillations}

\begin{figure}[h!]
\includegraphics[width=\columnwidth]{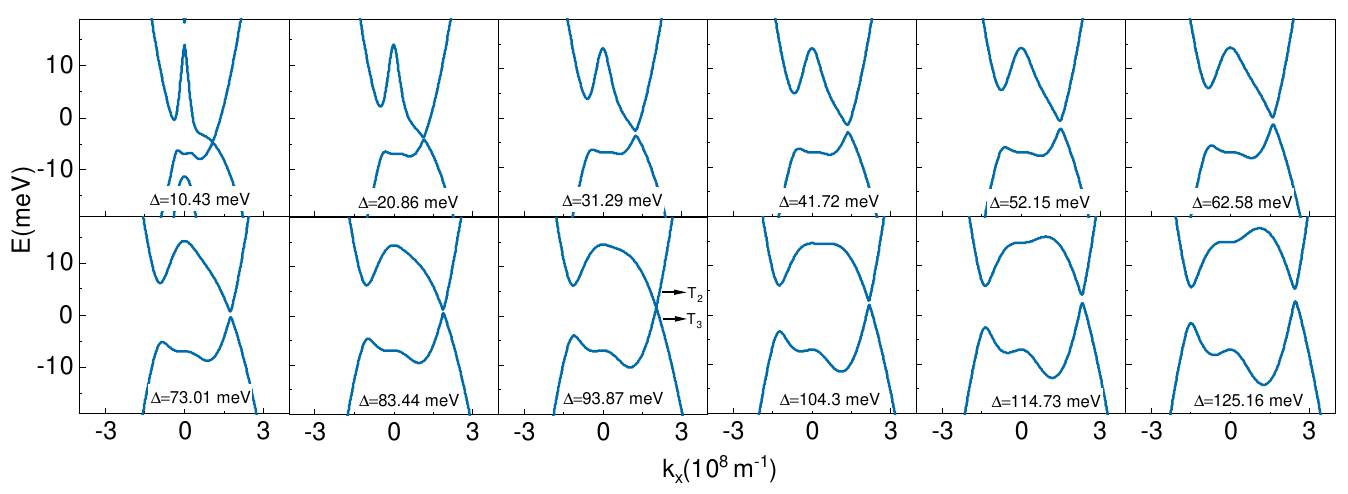}
\caption{\textbf{Calculation of Berry Phase}: Plots of the Landau level index $N$ (left axis) and $R_{xx}$ (right axis) as a
function of 1/B for (a) D = 0 V/nm (solid blue line), (b) D = 0.3 V/nm (solid green line) and (c) D = 1.1 V/nm (solid violet line). The scatter plots show the Landau level index($N$) vs $1/B$. The dotted red lines fit the $N$ versus $1/B$ data. The y-intercepts of the lines gives $\gamma$.}
\label{figS3}
\end{figure}

For a constant number density, $n$, and at a fixed $T$, the amplitude of the SdH oscillation is:
\begin{equation}
 \Delta R_{xx} \propto \frac{\lambda}{\mathrm{sinh} \lambda }  \mathrm{cos}\left( 2\pi\left( \frac{B_f}{B} +\gamma \right)\right)
 \label{eq_2}
 \end{equation}
\noindent The phase of oscillation is given by
 \begin{equation}
 \gamma=\frac{\Phi_B}{2\pi}-\frac{1}{2}
 \label{eq_3}
 \end{equation}
 The oscillations will have minima at
 \begin{equation}
    2\pi\left( \frac{B_f}{B} +\gamma \right)=(2N+1)\pi~~~~~~~~N=0, 1, 2..
 \label{eq_4}
 \end{equation}
or,
 \begin{equation}
    N=B_f\left(\frac{1}{B}\right)+\left(\gamma-\frac{1}{2}\right)
 \label{eq_5}
 \end{equation}
The intercept of the landau level index $N$ vs $1/B$ plot gives ($\gamma-1/2$).

Fig.~\ref{figS3}(a-c) show plots of $R_{xx}$ (right axis) and the Landau level index $N$ corresponding to minima in $R_{xx}$ (left axis) as a function of $1/B$ for $D=0~\mathrm{V/nm}$, $D=0.3~\mathrm{V/nm}$ and $D=1.1~\mathrm{V/nm}$. The y-axis intercept of the linear fit to the data (red dotted line in Fig.~\ref{figS3}(a~-~c)) gives $\Phi_B$ in units of $\pi$. For $D=0~\mathrm{V/nm}$, $\Phi_B = \pi$, this observation matches previous studies and is understood to originate from the underlying ${0^-_{M2}}$ band that adds two additional Landau levels to the sequence of BLL bands ~\cite{Nontrivialqhbeeryphase}. For $D=0.3~\mathrm{V/nm}$ and $D=1.1~\mathrm{V/nm}$, $\Phi_B =0$ showing that the gullies are massive.

\subsection{Evolution of $T_2$ and $T_3$ Gullies with displacement field $D$}

\begin{figure}[h!]
\includegraphics[width=\columnwidth]{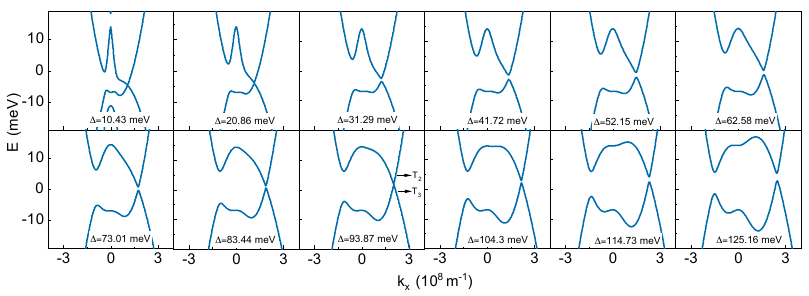}
\caption{\textbf{Evolution of Gully with $D$:}~ Energy $E$ vs $k_x$ for $D=0.1~\mathrm{V/nm}$ to $D=1.2~\mathrm{V/nm}$. The conversion factor is  $\Delta=104.3~\mathrm{meV/Vnm}$. }
\label{figS4}
\end{figure}

Fig.~\ref{figS4} shows the evolution of the $T_2$ and $T_3$ gullies with the displacement field. As the displacement field $D$ increases, the behavior of the $T_2$ and $T_3$ gullies becomes Dirac-like. At low $D$, the gully cones have a distorted parabolic shape, and as $D$  increases, the gully cones start becoming sharp like a monolayer graphene band, which we call a Dirac band. The band gap between the $T_2$ and $T_3$ gullies closes at $\Delta_c = 93.87$~meV. After $\Delta_c$, the band gap increases again with $D$.
It can also be seen that with increasing $D$, the inter-gully distance increases.

\subsection{Evolution of Landau level sequence and Berry phase with $n$ and $D$. }
\begin{figure}[h!]
\includegraphics[width=0.9\columnwidth]{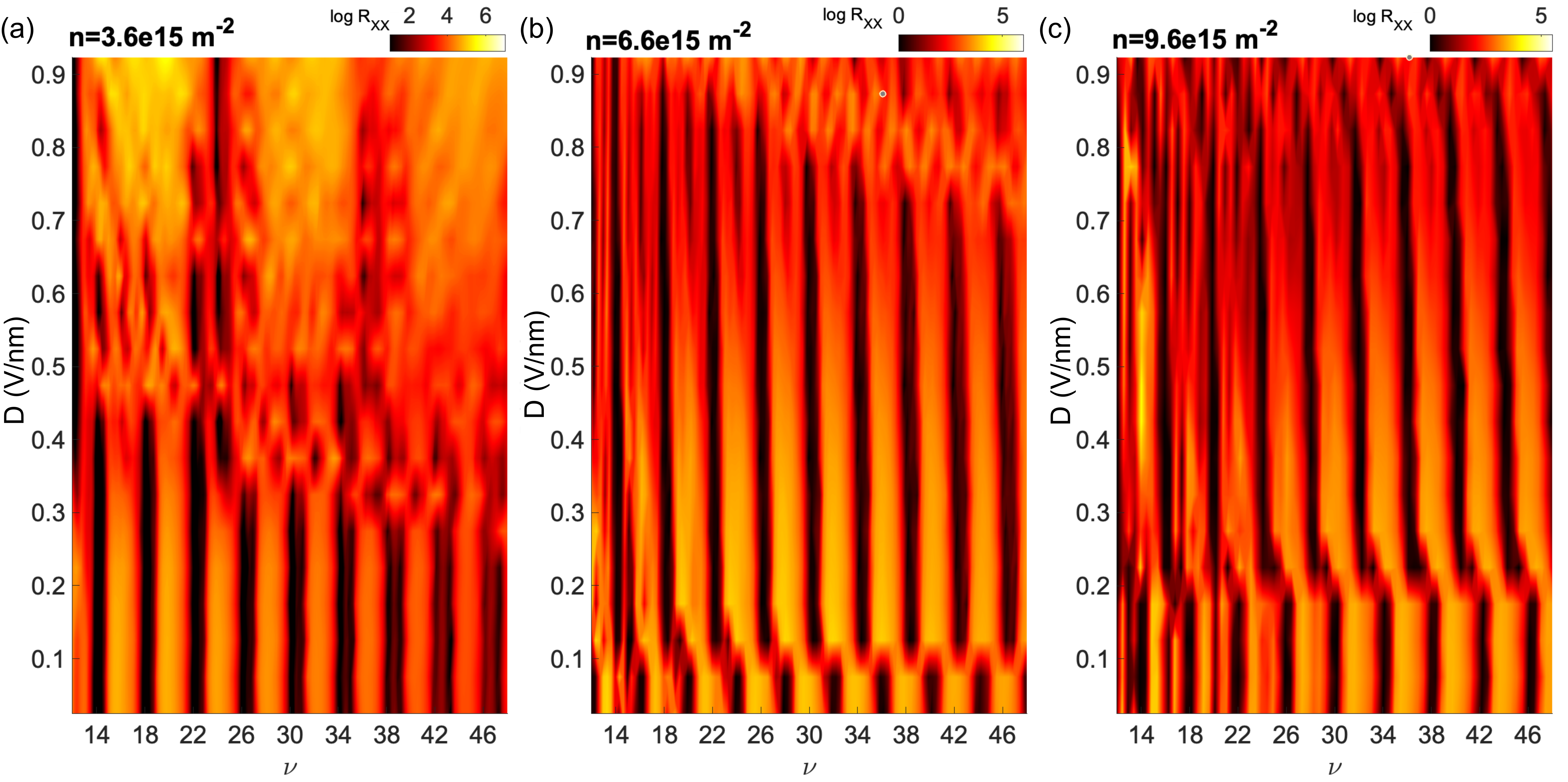}
\caption{\textbf{Evolution of Berry phase with $n$ and $D$:}~Contour map of $R_{xx}$ versus $D$ and $\nu$ at a constant number density (a) $n = 3.6\times 10^{15}~\mathrm{m^{-2}}$. (b) $n = 6.6\times 10^{15}~\mathrm{m^{-2}}$. (c) $n = 9.6\times 10^{15}~\mathrm{m^{-2}}$.}
\label{figS5}
\end{figure}

Fig.~\ref{figS5}(a-c) shows the contour maps of $R_{xx}$ vs $D$ for $n = 3.6\times 10^{15}~\mathrm{m^{-2}}$, $n = 6.6\times 10^{15}~\mathrm{m^{-2}}$ and $n = 9.6\times 10^{15}~\mathrm{m^{-2}}$. In Fig.~\ref{figS5}(a) at low $D$ a landau level sequence of $\nu=4N+2$ can be seen. The factor $2$ arises from the underlying zeroth Landau levels of the MLL band, which results in the berry phase of the BLL band to be   $\Phi_B=\pi$ \cite{Nontrivialqhbeeryphase}. For $D\geq0.5~\mathrm{V/nm}$, the emergence of landau level sequence $\nu=12N$ can be observed. This is due to the emergence of the three Gullies with 2-fold spin and 2-fold valley degeneracy. The Gullies in this region are massive as their Berry phase is $2\pi$.
In Fig.~\ref{figS5}(b) at low $D$ a landau level sequence of $\nu=4N$ with $\Phi_B=2\pi$ is observed. This is because the BLL band lies in the band gap of the MLL band in this region. For $D\geq0.1~\mathrm{V/nm}$ landau level sequence changes to $\nu=4N+2$ as the BLL band now lies in the valence band of the monolayer.
In Fig.~\ref{figS5}(c) at low $D$, a landau level sequence of $\nu=4(N+1/2)$ with $\Phi_B=\pi$ is observed, which is because the BLL band lies in the conduction band of monolayer band in this region. For $D\geq0.2~\mathrm{V/nm}$, the landau level sequence changes to $\nu=4N$ as the BLL band now lies in the band gap of the MLL band.


\begin{thebibliography}{37}%
	\makeatletter
	\providecommand \@ifxundefined [1]{%
		\@ifx{#1\undefined}
	}%
	\providecommand \@ifnum [1]{%
		\ifnum #1\expandafter \@firstoftwo
		\else \expandafter \@secondoftwo
		\fi
	}%
	\providecommand \@ifx [1]{%
		\ifx #1\expandafter \@firstoftwo
		\else \expandafter \@secondoftwo
		\fi
	}%
	\providecommand \natexlab [1]{#1}%
	\providecommand \enquote  [1]{``#1''}%
	\providecommand \bibnamefont  [1]{#1}%
	\providecommand \bibfnamefont [1]{#1}%
	\providecommand \citenamefont [1]{#1}%
	\providecommand \href@noop [0]{\@secondoftwo}%
	\providecommand \href [0]{\begingroup \@sanitize@url \@href}%
	\providecommand \@href[1]{\@@startlink{#1}\@@href}%
	\providecommand \@@href[1]{\endgroup#1\@@endlink}%
	\providecommand \@sanitize@url [0]{\catcode `\\12\catcode `\$12\catcode
		`\&12\catcode `\#12\catcode `\^12\catcode `\_12\catcode `\%12\relax}%
	\providecommand \@@startlink[1]{}%
	\providecommand \@@endlink[0]{}%
	\providecommand \url  [0]{\begingroup\@sanitize@url \@url }%
	\providecommand \@url [1]{\endgroup\@href {#1}{\urlprefix }}%
	\providecommand \urlprefix  [0]{URL }%
	\providecommand \Eprint [0]{\href }%
	\providecommand \doibase [0]{https://doi.org/}%
	\providecommand \selectlanguage [0]{\@gobble}%
	\providecommand \bibinfo  [0]{\@secondoftwo}%
	\providecommand \bibfield  [0]{\@secondoftwo}%
	\providecommand \translation [1]{[#1]}%
	\providecommand \BibitemOpen [0]{}%
	\providecommand \bibitemStop [0]{}%
	\providecommand \bibitemNoStop [0]{.\EOS\space}%
	\providecommand \EOS [0]{\spacefactor3000\relax}%
	\providecommand \BibitemShut  [1]{\csname bibitem#1\endcsname}%
	\let\auto@bib@innerbib\@empty
	\bibitem [{\citenamefont {Berry}(1984)}]{doi:10.1098/rspa.1984.0023}%
	\BibitemOpen
	\bibfield  {author} {\bibinfo {author} {\bibfnamefont {M.~V.}\ \bibnamefont
			{Berry}},\ }\bibfield  {title} {\bibinfo {title} {Quantal phase factors
			accompanying adiabatic changes},\ }\href
	{https://doi.org/10.1098/rspa.1984.0023} {\bibfield  {journal} {\bibinfo
			{journal} {Proceedings of the Royal Society of London. A. Mathematical and
				Physical Sciences}\ }\textbf {\bibinfo {volume} {392}},\ \bibinfo {pages}
		{45} (\bibinfo {year} {1984})}\BibitemShut {NoStop}%
	\bibitem [{\citenamefont {Xiao}\ \emph {et~al.}(2010)\citenamefont {Xiao},
		\citenamefont {Chang},\ and\ \citenamefont {Niu}}]{RevModPhys.82.1959}%
	\BibitemOpen
	\bibfield  {author} {\bibinfo {author} {\bibfnamefont {D.}~\bibnamefont
			{Xiao}}, \bibinfo {author} {\bibfnamefont {M.-C.}\ \bibnamefont {Chang}},\
		and\ \bibinfo {author} {\bibfnamefont {Q.}~\bibnamefont {Niu}},\ }\bibfield
	{title} {\bibinfo {title} {Berry phase effects on electronic properties},\
	}\href {https://doi.org/10.1103/RevModPhys.82.1959} {\bibfield  {journal}
		{\bibinfo  {journal} {Rev. Mod. Phys.}\ }\textbf {\bibinfo {volume} {82}},\
		\bibinfo {pages} {1959} (\bibinfo {year} {2010})}\BibitemShut {NoStop}%
	\bibitem [{\citenamefont {Mikitik}\ and\ \citenamefont
		{Sharlai}(1999)}]{Manisfestationofbp}%
	\BibitemOpen
	\bibfield  {author} {\bibinfo {author} {\bibfnamefont {G.~P.}\ \bibnamefont
			{Mikitik}}\ and\ \bibinfo {author} {\bibfnamefont {Y.~V.}\ \bibnamefont
			{Sharlai}},\ }\bibfield  {title} {\bibinfo {title} {Manifestation of berry's
			phase in metal physics},\ }\href
	{https://doi.org/10.1103/PhysRevLett.82.2147} {\bibfield  {journal} {\bibinfo
			{journal} {Phys. Rev. Lett.}\ }\textbf {\bibinfo {volume} {82}},\ \bibinfo
		{pages} {2147} (\bibinfo {year} {1999})}\BibitemShut {NoStop}%
	\bibitem [{\citenamefont {Luk'yanchuk}\ and\ \citenamefont
		{Kopelevich}(2004)}]{PhysRevLett.93.166402}%
	\BibitemOpen
	\bibfield  {author} {\bibinfo {author} {\bibfnamefont {I.~A.}\ \bibnamefont
			{Luk'yanchuk}}\ and\ \bibinfo {author} {\bibfnamefont {Y.}~\bibnamefont
			{Kopelevich}},\ }\bibfield  {title} {\bibinfo {title} {Phase analysis of
			quantum oscillations in graphite},\ }\href
	{https://doi.org/10.1103/PhysRevLett.93.166402} {\bibfield  {journal}
		{\bibinfo  {journal} {Phys. Rev. Lett.}\ }\textbf {\bibinfo {volume} {93}},\
		\bibinfo {pages} {166402} (\bibinfo {year} {2004})}\BibitemShut {NoStop}%
	\bibitem [{\citenamefont {Koshino}\ and\ \citenamefont
		{McCann}(2009)}]{2koshino_trigonal_2009}%
	\BibitemOpen
	\bibfield  {author} {\bibinfo {author} {\bibfnamefont {M.}~\bibnamefont
			{Koshino}}\ and\ \bibinfo {author} {\bibfnamefont {E.}~\bibnamefont
			{McCann}},\ }\bibfield  {title} {\bibinfo {title} {Trigonal warping and
			berry's phase $n\ensuremath{\pi}$ in abc-stacked multilayer graphene},\
	}\href {https://doi.org/10.1103/PhysRevB.80.165409} {\bibfield  {journal}
		{\bibinfo  {journal} {Phys. Rev. B}\ }\textbf {\bibinfo {volume} {80}},\
		\bibinfo {pages} {165409} (\bibinfo {year} {2009})}\BibitemShut {NoStop}%
	\bibitem [{\citenamefont {Zhang}\ \emph {et~al.}(2005)\citenamefont {Zhang},
		\citenamefont {Tan}, \citenamefont {Stormer},\ and\ \citenamefont
		{Kim}}]{3zhang_experimental_2005}%
	\BibitemOpen
	\bibfield  {author} {\bibinfo {author} {\bibfnamefont {Y.}~\bibnamefont
			{Zhang}}, \bibinfo {author} {\bibfnamefont {Y.-W.}\ \bibnamefont {Tan}},
		\bibinfo {author} {\bibfnamefont {H.~L.}\ \bibnamefont {Stormer}},\ and\
		\bibinfo {author} {\bibfnamefont {P.}~\bibnamefont {Kim}},\ }\bibfield
	{title} {\bibinfo {title} {Experimental observation of the quantum hall
			effect and berry's phase in graphene},\ }\href
	{https://doi.org/10.1038/nature04235} {\bibfield  {journal} {\bibinfo
			{journal} {Nature}\ }\textbf {\bibinfo {volume} {438}},\ \bibinfo {pages}
		{201} (\bibinfo {year} {2005})}\BibitemShut {NoStop}%
	\bibitem [{\citenamefont {Zhang}\ \emph {et~al.}(2011)\citenamefont {Zhang},
		\citenamefont {Zhang}, \citenamefont {Camacho}, \citenamefont {Khodas},\ and\
		\citenamefont {Zaliznyak}}]{4zhang_experimental_2011}%
	\BibitemOpen
	\bibfield  {author} {\bibinfo {author} {\bibfnamefont {L.}~\bibnamefont
			{Zhang}}, \bibinfo {author} {\bibfnamefont {Y.}~\bibnamefont {Zhang}},
		\bibinfo {author} {\bibfnamefont {J.}~\bibnamefont {Camacho}}, \bibinfo
		{author} {\bibfnamefont {M.}~\bibnamefont {Khodas}},\ and\ \bibinfo {author}
		{\bibfnamefont {I.}~\bibnamefont {Zaliznyak}},\ }\bibfield  {title} {\bibinfo
		{title} {The experimental observation of quantum hall effect of l=3 chiral
			quasiparticles in trilayer graphene},\ }\href
	{https://doi.org/10.1038/nphys2104} {\bibfield  {journal} {\bibinfo
			{journal} {Nature Physics}\ }\textbf {\bibinfo {volume} {7}},\ \bibinfo
		{pages} {953} (\bibinfo {year} {2011})}\BibitemShut {NoStop}%
	\bibitem [{\citenamefont {Novoselov}\ \emph {et~al.}(2006)\citenamefont
		{Novoselov}, \citenamefont {McCann}, \citenamefont {Morozov}, \citenamefont
		{Fal'ko}, \citenamefont {Katsnelson}, \citenamefont {Zeitler}, \citenamefont
		{Jiang}, \citenamefont {Schedin},\ and\ \citenamefont
		{Geim}}]{5novoselov_unconventional_2006}%
	\BibitemOpen
	\bibfield  {author} {\bibinfo {author} {\bibfnamefont {K.~S.}\ \bibnamefont
			{Novoselov}}, \bibinfo {author} {\bibfnamefont {E.}~\bibnamefont {McCann}},
		\bibinfo {author} {\bibfnamefont {S.~V.}\ \bibnamefont {Morozov}}, \bibinfo
		{author} {\bibfnamefont {V.~I.}\ \bibnamefont {Fal'ko}}, \bibinfo {author}
		{\bibfnamefont {M.~I.}\ \bibnamefont {Katsnelson}}, \bibinfo {author}
		{\bibfnamefont {U.}~\bibnamefont {Zeitler}}, \bibinfo {author} {\bibfnamefont
			{D.}~\bibnamefont {Jiang}}, \bibinfo {author} {\bibfnamefont
			{F.}~\bibnamefont {Schedin}},\ and\ \bibinfo {author} {\bibfnamefont {A.~K.}\
			\bibnamefont {Geim}},\ }\bibfield  {title} {\bibinfo {title} {Unconventional
			quantum hall effect and berry's phase of 2$\pi$ in bilayer graphene},\ }\href
	{https://doi.org/10.1038/nphys245} {\bibfield  {journal} {\bibinfo  {journal}
			{Nature Physics}\ }\textbf {\bibinfo {volume} {2}},\ \bibinfo {pages} {177}
		(\bibinfo {year} {2006})}\BibitemShut {NoStop}%
	\bibitem [{\citenamefont {Park}\ and\ \citenamefont
		{Marzari}(2011)}]{6park_berry_2011}%
	\BibitemOpen
	\bibfield  {author} {\bibinfo {author} {\bibfnamefont {C.-H.}\ \bibnamefont
			{Park}}\ and\ \bibinfo {author} {\bibfnamefont {N.}~\bibnamefont {Marzari}},\
	}\bibfield  {title} {\bibinfo {title} {Berry phase and pseudospin winding
			number in bilayer graphene},\ }\href
	{https://doi.org/10.1103/PhysRevB.84.205440} {\bibfield  {journal} {\bibinfo
			{journal} {Phys. Rev. B}\ }\textbf {\bibinfo {volume} {84}},\ \bibinfo
		{pages} {205440} (\bibinfo {year} {2011})}\BibitemShut {NoStop}%
	\bibitem [{\citenamefont {Mikitik}\ and\ \citenamefont
		{Sharlai}(2008)}]{7mikitik_electron_2008}%
	\BibitemOpen
	\bibfield  {author} {\bibinfo {author} {\bibfnamefont {G.~P.}\ \bibnamefont
			{Mikitik}}\ and\ \bibinfo {author} {\bibfnamefont {Y.~V.}\ \bibnamefont
			{Sharlai}},\ }\bibfield  {title} {\bibinfo {title} {Electron energy spectrum
			and the berry phase in a graphite bilayer},\ }\href
	{https://doi.org/10.1103/PhysRevB.77.113407} {\bibfield  {journal} {\bibinfo
			{journal} {Phys. Rev. B}\ }\textbf {\bibinfo {volume} {77}},\ \bibinfo
		{pages} {113407} (\bibinfo {year} {2008})}\BibitemShut {NoStop}%
	\bibitem [{\citenamefont {Li}\ \emph {et~al.}(2021)\citenamefont {Li},
		\citenamefont {Sanz}, \citenamefont {Merino-D{\'i}ez}, \citenamefont
		{Vilas-Varela}, \citenamefont {Garcia-Lekue}, \citenamefont {Corso},
		\citenamefont {de~Oteyza}, \citenamefont {Frederiksen}, \citenamefont
		{Pe{\~{n}}a},\ and\ \citenamefont {Pascual}}]{li_topological_2021}%
	\BibitemOpen
	\bibfield  {author} {\bibinfo {author} {\bibfnamefont {J.}~\bibnamefont
			{Li}}, \bibinfo {author} {\bibfnamefont {S.}~\bibnamefont {Sanz}}, \bibinfo
		{author} {\bibfnamefont {N.}~\bibnamefont {Merino-D{\'i}ez}}, \bibinfo
		{author} {\bibfnamefont {M.}~\bibnamefont {Vilas-Varela}}, \bibinfo {author}
		{\bibfnamefont {A.}~\bibnamefont {Garcia-Lekue}}, \bibinfo {author}
		{\bibfnamefont {M.}~\bibnamefont {Corso}}, \bibinfo {author} {\bibfnamefont
			{D.~G.}\ \bibnamefont {de~Oteyza}}, \bibinfo {author} {\bibfnamefont
			{T.}~\bibnamefont {Frederiksen}}, \bibinfo {author} {\bibfnamefont
			{D.}~\bibnamefont {Pe{\~{n}}a}},\ and\ \bibinfo {author} {\bibfnamefont
			{J.~I.}\ \bibnamefont {Pascual}},\ }\bibfield  {title} {\bibinfo {title}
		{Topological phase transition in chiral graphene nanoribbons: from edge bands
			to end states},\ }\href {https://doi.org/10.1038/s41467-021-25688-z}
	{\bibfield  {journal} {\bibinfo  {journal} {Nature Communications}\ }\textbf
		{\bibinfo {volume} {12}},\ \bibinfo {pages} {5538} (\bibinfo {year}
		{2021})}\BibitemShut {NoStop}%
	\bibitem [{\citenamefont {Liu}\ \emph {et~al.}(2021)\citenamefont {Liu},
		\citenamefont {Xu}, \citenamefont {Liu}, \citenamefont {Shen}, \citenamefont
		{Le}, \citenamefont {Li}, \citenamefont {Pei}, \citenamefont {Liang},
		\citenamefont {Dudin}, \citenamefont {Kim}, \citenamefont {Cacho},
		\citenamefont {Xu}, \citenamefont {Sun}, \citenamefont {Yang}, \citenamefont
		{Liu}, \citenamefont {Felser}, \citenamefont {Parkin},\ and\ \citenamefont
		{Chen}}]{PhysRevB.104.205140}%
	\BibitemOpen
	\bibfield  {author} {\bibinfo {author} {\bibfnamefont {D.~F.}\ \bibnamefont
			{Liu}}, \bibinfo {author} {\bibfnamefont {Q.~N.}\ \bibnamefont {Xu}},
		\bibinfo {author} {\bibfnamefont {E.~K.}\ \bibnamefont {Liu}}, \bibinfo
		{author} {\bibfnamefont {J.~L.}\ \bibnamefont {Shen}}, \bibinfo {author}
		{\bibfnamefont {C.~C.}\ \bibnamefont {Le}}, \bibinfo {author} {\bibfnamefont
			{Y.~W.}\ \bibnamefont {Li}}, \bibinfo {author} {\bibfnamefont
			{D.}~\bibnamefont {Pei}}, \bibinfo {author} {\bibfnamefont {A.~J.}\
			\bibnamefont {Liang}}, \bibinfo {author} {\bibfnamefont {P.}~\bibnamefont
			{Dudin}}, \bibinfo {author} {\bibfnamefont {T.~K.}\ \bibnamefont {Kim}},
		\bibinfo {author} {\bibfnamefont {C.}~\bibnamefont {Cacho}}, \bibinfo
		{author} {\bibfnamefont {Y.~F.}\ \bibnamefont {Xu}}, \bibinfo {author}
		{\bibfnamefont {Y.}~\bibnamefont {Sun}}, \bibinfo {author} {\bibfnamefont
			{L.~X.}\ \bibnamefont {Yang}}, \bibinfo {author} {\bibfnamefont {Z.~K.}\
			\bibnamefont {Liu}}, \bibinfo {author} {\bibfnamefont {C.}~\bibnamefont
			{Felser}}, \bibinfo {author} {\bibfnamefont {S.~S.~P.}\ \bibnamefont
			{Parkin}},\ and\ \bibinfo {author} {\bibfnamefont {Y.~L.}\ \bibnamefont
			{Chen}},\ }\bibfield  {title} {\bibinfo {title} {Topological phase transition
			in a magnetic weyl semimetal},\ }\href
	{https://doi.org/10.1103/PhysRevB.104.205140} {\bibfield  {journal} {\bibinfo
			{journal} {Phys. Rev. B}\ }\textbf {\bibinfo {volume} {104}},\ \bibinfo
		{pages} {205140} (\bibinfo {year} {2021})}\BibitemShut {NoStop}%
	\bibitem [{\citenamefont {König}\ \emph {et~al.}(2007)\citenamefont {König},
		\citenamefont {Wiedmann}, \citenamefont {Brüne}, \citenamefont {Roth},
		\citenamefont {Buhmann}, \citenamefont {Molenkamp}, \citenamefont {Qi},\ and\
		\citenamefont {Zhang}}]{doi:10.1126/science.1148047}%
	\BibitemOpen
	\bibfield  {author} {\bibinfo {author} {\bibfnamefont {M.}~\bibnamefont
			{König}}, \bibinfo {author} {\bibfnamefont {S.}~\bibnamefont {Wiedmann}},
		\bibinfo {author} {\bibfnamefont {C.}~\bibnamefont {Brüne}}, \bibinfo
		{author} {\bibfnamefont {A.}~\bibnamefont {Roth}}, \bibinfo {author}
		{\bibfnamefont {H.}~\bibnamefont {Buhmann}}, \bibinfo {author} {\bibfnamefont
			{L.~W.}\ \bibnamefont {Molenkamp}}, \bibinfo {author} {\bibfnamefont {X.-L.}\
			\bibnamefont {Qi}},\ and\ \bibinfo {author} {\bibfnamefont {S.-C.}\
			\bibnamefont {Zhang}},\ }\bibfield  {title} {\bibinfo {title} {Quantum spin
			hall insulator state in hgte quantum wells},\ }\href
	{https://doi.org/10.1126/science.1148047} {\bibfield  {journal} {\bibinfo
			{journal} {Science}\ }\textbf {\bibinfo {volume} {318}},\ \bibinfo {pages}
		{766} (\bibinfo {year} {2007})}\BibitemShut {NoStop}%
	\bibitem [{\citenamefont {Stepanov}\ \emph {et~al.}(2019)\citenamefont
		{Stepanov}, \citenamefont {Barlas}, \citenamefont {Che}, \citenamefont
		{Myhro}, \citenamefont {Voigt}, \citenamefont {Pi}, \citenamefont {Watanabe},
		\citenamefont {Taniguchi}, \citenamefont {Smirnov}, \citenamefont {Zhang},
		\citenamefont {Lake}, \citenamefont {MacDonald},\ and\ \citenamefont
		{Lau}}]{Quantumparityhall}%
	\BibitemOpen
	\bibfield  {author} {\bibinfo {author} {\bibfnamefont {P.}~\bibnamefont
			{Stepanov}}, \bibinfo {author} {\bibfnamefont {Y.}~\bibnamefont {Barlas}},
		\bibinfo {author} {\bibfnamefont {S.}~\bibnamefont {Che}}, \bibinfo {author}
		{\bibfnamefont {K.}~\bibnamefont {Myhro}}, \bibinfo {author} {\bibfnamefont
			{G.}~\bibnamefont {Voigt}}, \bibinfo {author} {\bibfnamefont
			{Z.}~\bibnamefont {Pi}}, \bibinfo {author} {\bibfnamefont {K.}~\bibnamefont
			{Watanabe}}, \bibinfo {author} {\bibfnamefont {T.}~\bibnamefont {Taniguchi}},
		\bibinfo {author} {\bibfnamefont {D.}~\bibnamefont {Smirnov}}, \bibinfo
		{author} {\bibfnamefont {F.}~\bibnamefont {Zhang}}, \bibinfo {author}
		{\bibfnamefont {R.~K.}\ \bibnamefont {Lake}}, \bibinfo {author}
		{\bibfnamefont {A.~H.}\ \bibnamefont {MacDonald}},\ and\ \bibinfo {author}
		{\bibfnamefont {C.~N.}\ \bibnamefont {Lau}},\ }\bibfield  {title} {\bibinfo
		{title} {Quantum parity hall effect in bernal-stacked trilayer graphene},\
	}\href {https://doi.org/10.1073/pnas.1820835116} {\bibfield  {journal}
		{\bibinfo  {journal} {Proceedings of the National Academy of Sciences}\
		}\textbf {\bibinfo {volume} {116}},\ \bibinfo {pages} {10286} (\bibinfo
		{year} {2019})}\BibitemShut {NoStop}%
	\bibitem [{\citenamefont {Huang}\ \emph {et~al.}(2024)\citenamefont {Huang},
		\citenamefont {Ke}, \citenamefont {Guan}, \citenamefont {Li},\ and\
		\citenamefont {Lou}}]{PhysRevB.109.045408}%
	\BibitemOpen
	\bibfield  {author} {\bibinfo {author} {\bibfnamefont {A.}~\bibnamefont
			{Huang}}, \bibinfo {author} {\bibfnamefont {S.}~\bibnamefont {Ke}}, \bibinfo
		{author} {\bibfnamefont {J.-H.}\ \bibnamefont {Guan}}, \bibinfo {author}
		{\bibfnamefont {J.}~\bibnamefont {Li}},\ and\ \bibinfo {author}
		{\bibfnamefont {W.-K.}\ \bibnamefont {Lou}},\ }\bibfield  {title} {\bibinfo
		{title} {Strain-induced topological phase transition in graphene
			nanoribbons},\ }\href {https://doi.org/10.1103/PhysRevB.109.045408}
	{\bibfield  {journal} {\bibinfo  {journal} {Phys. Rev. B}\ }\textbf {\bibinfo
			{volume} {109}},\ \bibinfo {pages} {045408} (\bibinfo {year}
		{2024})}\BibitemShut {NoStop}%
	\bibitem [{\citenamefont {Tiwari}\ \emph {et~al.}(2021)\citenamefont {Tiwari},
		\citenamefont {Srivastav},\ and\ \citenamefont
		{Bid}}]{PhysRevLett.126.096801}%
	\BibitemOpen
	\bibfield  {author} {\bibinfo {author} {\bibfnamefont {P.}~\bibnamefont
			{Tiwari}}, \bibinfo {author} {\bibfnamefont {S.~K.}\ \bibnamefont
			{Srivastav}},\ and\ \bibinfo {author} {\bibfnamefont {A.}~\bibnamefont
			{Bid}},\ }\bibfield  {title} {\bibinfo {title} {Electric-field-tunable valley
			zeeman effect in bilayer graphene heterostructures: Realization of the
			spin-orbit valve effect},\ }\href
	{https://doi.org/10.1103/PhysRevLett.126.096801} {\bibfield  {journal}
		{\bibinfo  {journal} {Phys. Rev. Lett.}\ }\textbf {\bibinfo {volume} {126}},\
		\bibinfo {pages} {096801} (\bibinfo {year} {2021})}\BibitemShut {NoStop}%
	\bibitem [{\citenamefont {Yao}\ \emph {et~al.}(2009)\citenamefont {Yao},
		\citenamefont {Yang},\ and\ \citenamefont {Niu}}]{PhysRevLett.102.096801}%
	\BibitemOpen
	\bibfield  {author} {\bibinfo {author} {\bibfnamefont {W.}~\bibnamefont
			{Yao}}, \bibinfo {author} {\bibfnamefont {S.~A.}\ \bibnamefont {Yang}},\ and\
		\bibinfo {author} {\bibfnamefont {Q.}~\bibnamefont {Niu}},\ }\bibfield
	{title} {\bibinfo {title} {Edge states in graphene: From gapped flat-band to
			gapless chiral modes},\ }\href
	{https://doi.org/10.1103/PhysRevLett.102.096801} {\bibfield  {journal}
		{\bibinfo  {journal} {Phys. Rev. Lett.}\ }\textbf {\bibinfo {volume} {102}},\
		\bibinfo {pages} {096801} (\bibinfo {year} {2009})}\BibitemShut {NoStop}%
	\bibitem [{\citenamefont {Tiwari}\ \emph {et~al.}(2022)\citenamefont {Tiwari},
		\citenamefont {Jat}, \citenamefont {Udupa}, \citenamefont {Narang},
		\citenamefont {Watanabe}, \citenamefont {Taniguchi}, \citenamefont {Sen},\
		and\ \citenamefont {Bid}}]{tiwari_experimental_2022}%
	\BibitemOpen
	\bibfield  {author} {\bibinfo {author} {\bibfnamefont {P.}~\bibnamefont
			{Tiwari}}, \bibinfo {author} {\bibfnamefont {M.~K.}\ \bibnamefont {Jat}},
		\bibinfo {author} {\bibfnamefont {A.}~\bibnamefont {Udupa}}, \bibinfo
		{author} {\bibfnamefont {D.~S.}\ \bibnamefont {Narang}}, \bibinfo {author}
		{\bibfnamefont {K.}~\bibnamefont {Watanabe}}, \bibinfo {author}
		{\bibfnamefont {T.}~\bibnamefont {Taniguchi}}, \bibinfo {author}
		{\bibfnamefont {D.}~\bibnamefont {Sen}},\ and\ \bibinfo {author}
		{\bibfnamefont {A.}~\bibnamefont {Bid}},\ }\bibfield  {title} {\bibinfo
		{title} {Experimental observation of spin  split energy dispersion in
			high  mobility single layer graphene/wse2 heterostructures},\ }\href
	{https://doi.org/10.1038/s41699-022-00348-y} {\bibfield  {journal} {\bibinfo
			{journal} {npj 2D Materials and Applications}\ }\textbf {\bibinfo {volume}
			{6}},\ \bibinfo {pages} {68} (\bibinfo {year} {2022})}\BibitemShut {NoStop}%
	\bibitem [{\citenamefont {Morimoto}\ and\ \citenamefont
		{Koshino}(2013)}]{gateinduceddiraccones}%
	\BibitemOpen
	\bibfield  {author} {\bibinfo {author} {\bibfnamefont {T.}~\bibnamefont
			{Morimoto}}\ and\ \bibinfo {author} {\bibfnamefont {M.}~\bibnamefont
			{Koshino}},\ }\bibfield  {title} {\bibinfo {title} {Gate-induced dirac cones
			in multilayer graphenes},\ }\href
	{https://doi.org/10.1103/PhysRevB.87.085424} {\bibfield  {journal} {\bibinfo
			{journal} {Phys. Rev. B}\ }\textbf {\bibinfo {volume} {87}},\ \bibinfo
		{pages} {085424} (\bibinfo {year} {2013})}\BibitemShut {NoStop}%
	\bibitem [{\citenamefont {Varlet}\ \emph {et~al.}(2014)\citenamefont {Varlet},
		\citenamefont {Bischoff}, \citenamefont {Simonet}, \citenamefont {Watanabe},
		\citenamefont {Taniguchi}, \citenamefont {Ihn}, \citenamefont {Ensslin},
		\citenamefont {Mucha-Kruczy\ifmmode~\acute{n}\else \'{n}\fi{}ski},\ and\
		\citenamefont {Fal'ko}}]{PhysRevLett.113.116602}%
	\BibitemOpen
	\bibfield  {author} {\bibinfo {author} {\bibfnamefont {A.}~\bibnamefont
			{Varlet}}, \bibinfo {author} {\bibfnamefont {D.}~\bibnamefont {Bischoff}},
		\bibinfo {author} {\bibfnamefont {P.}~\bibnamefont {Simonet}}, \bibinfo
		{author} {\bibfnamefont {K.}~\bibnamefont {Watanabe}}, \bibinfo {author}
		{\bibfnamefont {T.}~\bibnamefont {Taniguchi}}, \bibinfo {author}
		{\bibfnamefont {T.}~\bibnamefont {Ihn}}, \bibinfo {author} {\bibfnamefont
			{K.}~\bibnamefont {Ensslin}}, \bibinfo {author} {\bibfnamefont
			{M.}~\bibnamefont {Mucha-Kruczy\ifmmode~\acute{n}\else \'{n}\fi{}ski}},\ and\
		\bibinfo {author} {\bibfnamefont {V.~I.}\ \bibnamefont {Fal'ko}},\ }\bibfield
	{title} {\bibinfo {title} {Anomalous sequence of quantum hall liquids
			revealing a tunable lifshitz transition in bilayer graphene},\ }\href
	{https://doi.org/10.1103/PhysRevLett.113.116602} {\bibfield  {journal}
		{\bibinfo  {journal} {Phys. Rev. Lett.}\ }\textbf {\bibinfo {volume} {113}},\
		\bibinfo {pages} {116602} (\bibinfo {year} {2014})}\BibitemShut {NoStop}%
	\bibitem [{\citenamefont {Predin}\ \emph {et~al.}(2016)\citenamefont {Predin},
		\citenamefont {Wenk},\ and\ \citenamefont {Schliemann}}]{PhysRevB.93.115106}%
	\BibitemOpen
	\bibfield  {author} {\bibinfo {author} {\bibfnamefont {S.}~\bibnamefont
			{Predin}}, \bibinfo {author} {\bibfnamefont {P.}~\bibnamefont {Wenk}},\ and\
		\bibinfo {author} {\bibfnamefont {J.}~\bibnamefont {Schliemann}},\ }\bibfield
	{title} {\bibinfo {title} {Trigonal warping in bilayer graphene: Energy
			versus entanglement spectrum},\ }\href
	{https://doi.org/10.1103/PhysRevB.93.115106} {\bibfield  {journal} {\bibinfo
			{journal} {Phys. Rev. B}\ }\textbf {\bibinfo {volume} {93}},\ \bibinfo
		{pages} {115106} (\bibinfo {year} {2016})}\BibitemShut {NoStop}%
	\bibitem [{\citenamefont {Serbyn}\ and\ \citenamefont
		{Abanin}(2013)}]{newdiracpoints}%
	\BibitemOpen
	\bibfield  {author} {\bibinfo {author} {\bibfnamefont {M.}~\bibnamefont
			{Serbyn}}\ and\ \bibinfo {author} {\bibfnamefont {D.~A.}\ \bibnamefont
			{Abanin}},\ }\bibfield  {title} {\bibinfo {title} {New dirac points and
			multiple landau level crossings in biased trilayer graphene},\ }\href
	{https://doi.org/10.1103/PhysRevB.87.115422} {\bibfield  {journal} {\bibinfo
			{journal} {Phys. Rev. B}\ }\textbf {\bibinfo {volume} {87}},\ \bibinfo
		{pages} {115422} (\bibinfo {year} {2013})}\BibitemShut {NoStop}%
	\bibitem [{\citenamefont {Rao}\ and\ \citenamefont
		{Serbyn}(2020)}]{GullyQHferro}%
	\BibitemOpen
	\bibfield  {author} {\bibinfo {author} {\bibfnamefont {P.}~\bibnamefont
			{Rao}}\ and\ \bibinfo {author} {\bibfnamefont {M.}~\bibnamefont {Serbyn}},\
	}\bibfield  {title} {\bibinfo {title} {Gully quantum hall ferromagnetism in
			biased trilayer graphene},\ }\href
	{https://doi.org/10.1103/PhysRevB.101.245411} {\bibfield  {journal} {\bibinfo
			{journal} {Phys. Rev. B}\ }\textbf {\bibinfo {volume} {101}},\ \bibinfo
		{pages} {245411} (\bibinfo {year} {2020})}\BibitemShut {NoStop}%
	\bibitem [{\citenamefont {de~Gail}\ \emph {et~al.}(2012)\citenamefont
		{de~Gail}, \citenamefont {Goerbig},\ and\ \citenamefont
		{Montambaux}}]{PhysRevB.86.045407}%
	\BibitemOpen
	\bibfield  {author} {\bibinfo {author} {\bibfnamefont {R.}~\bibnamefont
			{de~Gail}}, \bibinfo {author} {\bibfnamefont {M.~O.}\ \bibnamefont
			{Goerbig}},\ and\ \bibinfo {author} {\bibfnamefont {G.}~\bibnamefont
			{Montambaux}},\ }\bibfield  {title} {\bibinfo {title} {Magnetic spectrum of
			trigonally warped bilayer graphene: Semiclassical analysis, zero modes, and
			topological winding numbers},\ }\href
	{https://doi.org/10.1103/PhysRevB.86.045407} {\bibfield  {journal} {\bibinfo
			{journal} {Phys. Rev. B}\ }\textbf {\bibinfo {volume} {86}},\ \bibinfo
		{pages} {045407} (\bibinfo {year} {2012})}\BibitemShut {NoStop}%
	\bibitem [{\citenamefont {Varlet}\ \emph {et~al.}(2015)\citenamefont {Varlet},
		\citenamefont {Mucha-Kruczynski}, \citenamefont {Bischoff}, \citenamefont
		{Simonet}, \citenamefont {Taniguchi}, \citenamefont {Watanabe}, \citenamefont
		{Falko}, \citenamefont {Ihn},\ and\ \citenamefont
		{Ensslin}}]{VARLET201519}%
	\BibitemOpen
	\bibfield  {author} {\bibinfo {author} {\bibfnamefont {A.}~\bibnamefont
			{Varlet}}, \bibinfo {author} {\bibfnamefont {M.}~\bibnamefont
			{Mucha-Kruczyński}}, \bibinfo {author} {\bibfnamefont {D.}~\bibnamefont
			{Bischoff}}, \bibinfo {author} {\bibfnamefont {P.}~\bibnamefont {Simonet}},
		\bibinfo {author} {\bibfnamefont {T.}~\bibnamefont {Taniguchi}}, \bibinfo
		{author} {\bibfnamefont {K.}~\bibnamefont {Watanabe}}, \bibinfo {author}
		{\bibfnamefont {V.}~\bibnamefont {Fal’ko}}, \bibinfo {author}
		{\bibfnamefont {T.}~\bibnamefont {Ihn}},\ and\ \bibinfo {author}
		{\bibfnamefont {K.}~\bibnamefont {Ensslin}},\ }\bibfield  {title} {\bibinfo
		{title} {Tunable fermi surface topology and lifshitz transition in bilayer
			graphene},\ }\href
	{https://doi.org/https://doi.org/10.1016/j.synthmet.2015.07.006} {\bibfield
		{journal} {\bibinfo  {journal} {Synthetic Metals}\ }\textbf {\bibinfo
			{volume} {210}},\ \bibinfo {pages} {19} (\bibinfo {year} {2015})},\ \bibinfo
	{note} {reviews of Current Advances in Graphene Science and
		Technology}\BibitemShut {NoStop}%
	\bibitem [{\citenamefont {Seiler}\ \emph {et~al.}(2022)\citenamefont {Seiler},
		\citenamefont {Geisenhof}, \citenamefont {Winterer}, \citenamefont
		{Watanabe}, \citenamefont {Taniguchi}, \citenamefont {Xu}, \citenamefont
		{Zhang},\ and\ \citenamefont {Weitz}}]{Seiler2022}%
	\BibitemOpen
	\bibfield  {author} {\bibinfo {author} {\bibfnamefont {A.~M.}\ \bibnamefont
			{Seiler}}, \bibinfo {author} {\bibfnamefont {F.~R.}\ \bibnamefont
			{Geisenhof}}, \bibinfo {author} {\bibfnamefont {F.}~\bibnamefont {Winterer}},
		\bibinfo {author} {\bibfnamefont {K.}~\bibnamefont {Watanabe}}, \bibinfo
		{author} {\bibfnamefont {T.}~\bibnamefont {Taniguchi}}, \bibinfo {author}
		{\bibfnamefont {T.}~\bibnamefont {Xu}}, \bibinfo {author} {\bibfnamefont
			{F.}~\bibnamefont {Zhang}},\ and\ \bibinfo {author} {\bibfnamefont {R.~T.}\
			\bibnamefont {Weitz}},\ }\bibfield  {title} {\bibinfo {title} {Quantum
			cascade of correlated phases in trigonally warped bilayer graphene},\ }\href
	{https://doi.org/10.1038/s41586-022-04937-1} {\bibfield  {journal} {\bibinfo
			{journal} {Nature}\ }\textbf {\bibinfo {volume} {608}},\ \bibinfo {pages}
		{298} (\bibinfo {year} {2022})}\BibitemShut {NoStop}%
	\bibitem [{\citenamefont {Qin}\ \emph {et~al.}(2025)\citenamefont {Qin},
		\citenamefont {Jing}, \citenamefont {Hao}, \citenamefont {Jiang},
		\citenamefont {Zhang}, \citenamefont {Cao}, \citenamefont {Song},\ and\
		\citenamefont {Guo}}]{PhysRevLett.134.036301}%
	\BibitemOpen
	\bibfield  {author} {\bibinfo {author} {\bibfnamefont {G.-Q.}\ \bibnamefont
			{Qin}}, \bibinfo {author} {\bibfnamefont {F.-M.}\ \bibnamefont {Jing}},
		\bibinfo {author} {\bibfnamefont {T.-Y.}\ \bibnamefont {Hao}}, \bibinfo
		{author} {\bibfnamefont {S.-L.}\ \bibnamefont {Jiang}}, \bibinfo {author}
		{\bibfnamefont {Z.-Z.}\ \bibnamefont {Zhang}}, \bibinfo {author}
		{\bibfnamefont {G.}~\bibnamefont {Cao}}, \bibinfo {author} {\bibfnamefont
			{X.-X.}\ \bibnamefont {Song}},\ and\ \bibinfo {author} {\bibfnamefont
			{G.-P.}\ \bibnamefont {Guo}},\ }\bibfield  {title} {\bibinfo {title}
		{Switching spin filling sequence in a bilayer graphene quantum dot through
			trigonal warping},\ }\href {https://doi.org/10.1103/PhysRevLett.134.036301}
	{\bibfield  {journal} {\bibinfo  {journal} {Phys. Rev. Lett.}\ }\textbf
		{\bibinfo {volume} {134}},\ \bibinfo {pages} {036301} (\bibinfo {year}
		{2025})}\BibitemShut {NoStop}%
	\bibitem [{\citenamefont {Zibrov}\ \emph {et~al.}(2018)\citenamefont {Zibrov},
		\citenamefont {Rao}, \citenamefont {Kometter}, \citenamefont {Spanton},
		\citenamefont {Li}, \citenamefont {Dean}, \citenamefont {Taniguchi},
		\citenamefont {Watanabe}, \citenamefont {Serbyn},\ and\ \citenamefont
		{Young}}]{emergentdiracgullies}%
	\BibitemOpen
	\bibfield  {author} {\bibinfo {author} {\bibfnamefont {A.~A.}\ \bibnamefont
			{Zibrov}}, \bibinfo {author} {\bibfnamefont {P.}~\bibnamefont {Rao}},
		\bibinfo {author} {\bibfnamefont {C.}~\bibnamefont {Kometter}}, \bibinfo
		{author} {\bibfnamefont {E.~M.}\ \bibnamefont {Spanton}}, \bibinfo {author}
		{\bibfnamefont {J.~I.~A.}\ \bibnamefont {Li}}, \bibinfo {author}
		{\bibfnamefont {C.~R.}\ \bibnamefont {Dean}}, \bibinfo {author}
		{\bibfnamefont {T.}~\bibnamefont {Taniguchi}}, \bibinfo {author}
		{\bibfnamefont {K.}~\bibnamefont {Watanabe}}, \bibinfo {author}
		{\bibfnamefont {M.}~\bibnamefont {Serbyn}},\ and\ \bibinfo {author}
		{\bibfnamefont {A.~F.}\ \bibnamefont {Young}},\ }\bibfield  {title} {\bibinfo
		{title} {Emergent dirac gullies and gully-symmetry-breaking quantum hall
			states in $aba$ trilayer graphene},\ }\href
	{https://doi.org/10.1103/PhysRevLett.121.167601} {\bibfield  {journal}
		{\bibinfo  {journal} {Phys. Rev. Lett.}\ }\textbf {\bibinfo {volume} {121}},\
		\bibinfo {pages} {167601} (\bibinfo {year} {2018})}\BibitemShut {NoStop}%
	\bibitem [{\citenamefont {Zhou}\ \emph {et~al.}(2023)\citenamefont {Zhou},
		\citenamefont {Auerbach}, \citenamefont {Uzan}, \citenamefont {Zhou},
		\citenamefont {Banu}, \citenamefont {Zhi}, \citenamefont {Huber},
		\citenamefont {Watanabe}, \citenamefont {Taniguchi}, \citenamefont
		{Myasoedov}, \citenamefont {Yan},\ and\ \citenamefont
		{Zeldov}}]{imagingquantumoscillations}%
	\BibitemOpen
	\bibfield  {author} {\bibinfo {author} {\bibfnamefont {H.}~\bibnamefont
			{Zhou}}, \bibinfo {author} {\bibfnamefont {N.}~\bibnamefont {Auerbach}},
		\bibinfo {author} {\bibfnamefont {M.}~\bibnamefont {Uzan}}, \bibinfo {author}
		{\bibfnamefont {Y.}~\bibnamefont {Zhou}}, \bibinfo {author} {\bibfnamefont
			{N.}~\bibnamefont {Banu}}, \bibinfo {author} {\bibfnamefont {W.}~\bibnamefont
			{Zhi}}, \bibinfo {author} {\bibfnamefont {M.~E.}\ \bibnamefont {Huber}},
		\bibinfo {author} {\bibfnamefont {K.}~\bibnamefont {Watanabe}}, \bibinfo
		{author} {\bibfnamefont {T.}~\bibnamefont {Taniguchi}}, \bibinfo {author}
		{\bibfnamefont {Y.}~\bibnamefont {Myasoedov}}, \bibinfo {author}
		{\bibfnamefont {B.}~\bibnamefont {Yan}},\ and\ \bibinfo {author}
		{\bibfnamefont {E.}~\bibnamefont {Zeldov}},\ }\bibfield  {title} {\bibinfo
		{title} {Imaging quantum oscillations and millitesla pseudomagnetic fields in
			graphene},\ }\href {https://doi.org/10.1038/s41586-023-06763-5} {\bibfield
		{journal} {\bibinfo  {journal} {Nature}\ }\textbf {\bibinfo {volume} {624}},\
		\bibinfo {pages} {275} (\bibinfo {year} {2023})}\BibitemShut {NoStop}%
	\bibitem [{\citenamefont {Srivastav}\ \emph {et~al.}(2024)\citenamefont
		{Srivastav}, \citenamefont {Udupa}, \citenamefont {Watanabe}, \citenamefont
		{Taniguchi}, \citenamefont {Sen},\ and\ \citenamefont {Das}}]{Nonlocaltlg}%
	\BibitemOpen
	\bibfield  {author} {\bibinfo {author} {\bibfnamefont {S.~K.}\ \bibnamefont
			{Srivastav}}, \bibinfo {author} {\bibfnamefont {A.}~\bibnamefont {Udupa}},
		\bibinfo {author} {\bibfnamefont {K.}~\bibnamefont {Watanabe}}, \bibinfo
		{author} {\bibfnamefont {T.}~\bibnamefont {Taniguchi}}, \bibinfo {author}
		{\bibfnamefont {D.}~\bibnamefont {Sen}},\ and\ \bibinfo {author}
		{\bibfnamefont {A.}~\bibnamefont {Das}},\ }\bibfield  {title} {\bibinfo
		{title} {Electric-field-tunable edge transport in bernal-stacked trilayer
			graphene},\ }\href {https://doi.org/10.1103/PhysRevLett.132.096301}
	{\bibfield  {journal} {\bibinfo  {journal} {Phys. Rev. Lett.}\ }\textbf
		{\bibinfo {volume} {132}},\ \bibinfo {pages} {096301} (\bibinfo {year}
		{2024})}\BibitemShut {NoStop}%
	\bibitem [{\citenamefont {Datta}\ \emph {et~al.}(2019)\citenamefont {Datta},
		\citenamefont {Adak}, \citenamefont {kun Shi}, \citenamefont {Watanabe},
		\citenamefont {Taniguchi}, \citenamefont {Song},\ and\ \citenamefont
		{Deshmukh}}]{Nontrivialqhbeeryphase}%
	\BibitemOpen
	\bibfield  {author} {\bibinfo {author} {\bibfnamefont {B.}~\bibnamefont
			{Datta}}, \bibinfo {author} {\bibfnamefont {P.~C.}\ \bibnamefont {Adak}},
		\bibinfo {author} {\bibfnamefont {L.}~\bibnamefont {kun Shi}}, \bibinfo
		{author} {\bibfnamefont {K.}~\bibnamefont {Watanabe}}, \bibinfo {author}
		{\bibfnamefont {T.}~\bibnamefont {Taniguchi}}, \bibinfo {author}
		{\bibfnamefont {J.~C.~W.}\ \bibnamefont {Song}},\ and\ \bibinfo {author}
		{\bibfnamefont {M.~M.}\ \bibnamefont {Deshmukh}},\ }\bibfield  {title}
	{\bibinfo {title} {Nontrivial quantum oscillation geometric phase shift in a
			trivial band},\ }\href {https://doi.org/10.1126/sciadv.aax6550} {\bibfield
		{journal} {\bibinfo  {journal} {Science Advances}\ }\textbf {\bibinfo
			{volume} {5}},\ \bibinfo {pages} {eaax6550} (\bibinfo {year}
		{2019})}\BibitemShut {NoStop}%
	\bibitem [{\citenamefont {Kaur}\ \emph {et~al.}(2024)\citenamefont {Kaur},
		\citenamefont {Chanda}, \citenamefont {Amin}, \citenamefont {Sahani},
		\citenamefont {Watanabe}, \citenamefont {Taniguchi}, \citenamefont {Ghorai},
		\citenamefont {Gefen}, \citenamefont {Sreejith},\ and\ \citenamefont
		{Bid}}]{kaur_universality_2024}%
	\BibitemOpen
	\bibfield  {author} {\bibinfo {author} {\bibfnamefont {S.}~\bibnamefont
			{Kaur}}, \bibinfo {author} {\bibfnamefont {T.}~\bibnamefont {Chanda}},
		\bibinfo {author} {\bibfnamefont {K.~R.}\ \bibnamefont {Amin}}, \bibinfo
		{author} {\bibfnamefont {D.}~\bibnamefont {Sahani}}, \bibinfo {author}
		{\bibfnamefont {K.}~\bibnamefont {Watanabe}}, \bibinfo {author}
		{\bibfnamefont {T.}~\bibnamefont {Taniguchi}}, \bibinfo {author}
		{\bibfnamefont {U.}~\bibnamefont {Ghorai}}, \bibinfo {author} {\bibfnamefont
			{Y.}~\bibnamefont {Gefen}}, \bibinfo {author} {\bibfnamefont {G.~J.}\
			\bibnamefont {Sreejith}},\ and\ \bibinfo {author} {\bibfnamefont
			{A.}~\bibnamefont {Bid}},\ }\bibfield  {title} {\bibinfo {title}
		{Universality of quantum phase transitions in the integer and fractional
			quantum hall regimes},\ }\href {https://doi.org/10.1038/s41467-024-52927-w}
	{\bibfield  {journal} {\bibinfo  {journal} {Nature Communications}\ }\textbf
		{\bibinfo {volume} {15}},\ \bibinfo {pages} {8535} (\bibinfo {year}
		{2024})}\BibitemShut {NoStop}%
	\bibitem [{\citenamefont {K\"uppersbusch}\ and\ \citenamefont
		{Fritz}(2017)}]{PhysRevB.96.205410}%
	\BibitemOpen
	\bibfield  {author} {\bibinfo {author} {\bibfnamefont {C.}~\bibnamefont
			{K\"uppersbusch}}\ and\ \bibinfo {author} {\bibfnamefont {L.}~\bibnamefont
			{Fritz}},\ }\bibfield  {title} {\bibinfo {title} {Modifications of the
			lifshitz-kosevich formula in two-dimensional dirac systems},\ }\href
	{https://doi.org/10.1103/PhysRevB.96.205410} {\bibfield  {journal} {\bibinfo
			{journal} {Phys. Rev. B}\ }\textbf {\bibinfo {volume} {96}},\ \bibinfo
		{pages} {205410} (\bibinfo {year} {2017})}\BibitemShut {NoStop}%
	\bibitem [{\citenamefont {Novoselov}\ \emph {et~al.}(2005)\citenamefont
		{Novoselov}, \citenamefont {Geim}, \citenamefont {Morozov}, \citenamefont
		{Jiang}, \citenamefont {Katsnelson}, \citenamefont {Grigorieva},
		\citenamefont {Dubonos},\ and\ \citenamefont
		{Firsov}}]{1novoselov_two-dimensional_2005}%
	\BibitemOpen
	\bibfield  {author} {\bibinfo {author} {\bibfnamefont {K.~S.}\ \bibnamefont
			{Novoselov}}, \bibinfo {author} {\bibfnamefont {A.~K.}\ \bibnamefont {Geim}},
		\bibinfo {author} {\bibfnamefont {S.~V.}\ \bibnamefont {Morozov}}, \bibinfo
		{author} {\bibfnamefont {D.}~\bibnamefont {Jiang}}, \bibinfo {author}
		{\bibfnamefont {M.~I.}\ \bibnamefont {Katsnelson}}, \bibinfo {author}
		{\bibfnamefont {I.~V.}\ \bibnamefont {Grigorieva}}, \bibinfo {author}
		{\bibfnamefont {S.~V.}\ \bibnamefont {Dubonos}},\ and\ \bibinfo {author}
		{\bibfnamefont {A.~A.}\ \bibnamefont {Firsov}},\ }\bibfield  {title}
	{\bibinfo {title} {Two-dimensional gas of massless dirac fermions in
			graphene},\ }\href {https://doi.org/10.1038/nature04233} {\bibfield
		{journal} {\bibinfo  {journal} {Nature}\ }\textbf {\bibinfo {volume} {438}},\
		\bibinfo {pages} {197} (\bibinfo {year} {2005})}\BibitemShut {NoStop}%
	\bibitem [{\citenamefont {Dresselhaus}\ and\ \citenamefont
		{Dresselhaus}(2002)}]{Dresselhaus01012002}%
	\BibitemOpen
	\bibfield  {author} {\bibinfo {author} {\bibfnamefont {M.~S.}\ \bibnamefont
			{Dresselhaus}}\ and\ \bibinfo {author} {\bibfnamefont {G.}~\bibnamefont
			{Dresselhaus}},\ }\bibfield  {title} {\bibinfo {title} {Intercalation
			compounds of graphite},\ }\href {https://doi.org/10.1080/00018730110113644}
	{\bibfield  {journal} {\bibinfo  {journal} {Advances in Physics}\ }\textbf
		{\bibinfo {volume} {51}},\ \bibinfo {pages} {1} (\bibinfo {year}
		{2002})}\BibitemShut {NoStop}%
	\bibitem [{\citenamefont {Slonczewski}\ and\ \citenamefont
		{Weiss}(1958)}]{PhysRev.109.272}%
	\BibitemOpen
	\bibfield  {author} {\bibinfo {author} {\bibfnamefont {J.~C.}\ \bibnamefont
			{Slonczewski}}\ and\ \bibinfo {author} {\bibfnamefont {P.~R.}\ \bibnamefont
			{Weiss}},\ }\bibfield  {title} {\bibinfo {title} {Band structure of
			graphite},\ }\href {https://doi.org/10.1103/PhysRev.109.272} {\bibfield
		{journal} {\bibinfo  {journal} {Phys. Rev.}\ }\textbf {\bibinfo {volume}
			{109}},\ \bibinfo {pages} {272} (\bibinfo {year} {1958})}\BibitemShut
	{NoStop}%
	\bibitem [{\citenamefont {McClure}(1969)}]{MCCLURE1969425}%
	\BibitemOpen
	\bibfield  {author} {\bibinfo {author} {\bibfnamefont {J.}~\bibnamefont
			{McClure}},\ }\bibfield  {title} {\bibinfo {title} {Electron energy band
			structure and electronic properties of rhombohedral graphite},\ }\href
	{https://doi.org/https://doi.org/10.1016/0008-6223(69)90073-6} {\bibfield
		{journal} {\bibinfo  {journal} {Carbon}\ }\textbf {\bibinfo {volume} {7}},\
		\bibinfo {pages} {425} (\bibinfo {year} {1969})}\BibitemShut {NoStop}%
\end{thebibliography}

%

\end{document}